\documentclass[bibyear]{aa} 
\usepackage{txfonts}
\usepackage{upgreek}
\usepackage{tipa}
\usepackage{verbatim}
\usepackage{natbib}
\usepackage{color}
\usepackage[hang,small]{caption}

\begin{document}

   \title{Magnetic field in IRC+10216 and other C-Rich Evolved Stars}


   \author{A. Duthu
          \inst{1}
          \and
          F. Herpin\inst{1}
           \and
          H. Wiesemeyer\inst{2}
          \and
          A. Baudry\inst{1}
           \and
          A. L\`{e}bre\inst{3}
            \and
        G. Paubert\inst{4}
          }

   \institute{
   Laboratoire d'astrophysique de Bordeaux, Univ. Bordeaux, CNRS, B18N, all\'ee Geoffroy Saint-Hilaire, F-33615 Pessac, France
              \email{alizee.duthu@u-bordeaux.fr}
    \and 
    Max-Planck-Institut f\"ur Radioastronomie, Auf dem H\"ugel 69, 53121 Bonn, Germany
    \and
    Laboratoire Univers et Particules de Montpellier, CNRS et Universit\'e de Montpellier, Place E. Bataillon, 34090 Montpellier, France
        \and
    Instituto Radioastronom\'{i}a Milim\'etrica (IRAM), Av. Divina Pastora 7, N\'ucleo Central, 18012 Granada, Spain
}

 
  \abstract
   { During the transition from the Asymptotic Giant Branch (AGB) to Planetary Nebulae (PN), the circumstellar geometry and morphology change dramatically. Another characteristic of this transition is the high mass loss rate, that can be partially explained by radiation pressure and a combination of various factors like the stellar pulsation, the dust grain condensation and opacity in the upper atmosphere. The magnetic field can also be one of the main ingredients that shapes the stellar upper atmosphere and envelope.}
   {Our main goal is to investigate for the first time the spatial distribution of the magnetic field in the envelope of IRC+10216. More generally we intend to determine the magnetic field strength in the circumstellar envelope (CSE) of C-rich evolved stars, compare this field with previous studies for O-rich stars, and constrain the variation of the magnetic field with r the distance to the star's center.
}
   {We use spectropolarimetric observations of the Stokes $V$ parameter, collected with Xpol on the IRAM-30m radiotelescope, observing the Zeeman effect in seven hyperfine components of the CN J = 1-0 line. We use Crutcher's method to estimate the magnetic field.
For the first time, the instrumental contamination is investigated, through dedicated studies of the power patterns in Stokes \textit{V} and \textit{I} in detail.
}
   {For C-rich evolved stars, we derive a magnetic field strength (B) between 1.6 and 14.2 mG while B is estimated to be 6 mG for the proto-PN (PPN) AFGL618, and an upper value of 8 mG is found for the PN NGC7027. These results are consistent with a decrease of B as 1/r in  the environment of AGB objects, i.e., with the presence of a toroidal field. But this is not the case for PPN and PN stars. Our map of IRC+10216 suggests that the magnetic field is not homogeneously strong throughout or aligned with the envelope and that the morphology of the CN emission might have changed with time. }
   {}


   \maketitle
%

\section{Introduction}


Prior to the Planetary Nebula (PN) stage, stars ascend the Asymptotic Giant Branch (AGB) phase and become thousands of times more luminous than on the main sequence. The central object experiences thermal pulses \citep[][]{habing1996}. 
In addition, these stars loose significant amounts of their mass in the form of stellar winds. These outflows will form chemically rich circumstellar envelopes (CSEs) around the AGB stars. The more massive AGB stars exhibiting larger luminosities, longer pulsation periods, enhanced mass losses, also show strong OH maser emission and are called OH$/$IR stars. This transition from AGB to PN is hence characterized by a high mass loss rate \citep[e.g.,][]{debeck2010}. The mass loss mechanism is driven mainly by the radiation pressure on the dust although a combination of several other factors may also play an important role \citep[e.g.,][]{hofner2016}, including the stellar pulsation, the condensation and opacity of dust stellar grains in the upper atmosphere, and magnetic activity. 

\begin{table*}
\begin{small}
\caption{Source list and observation parameters. The rms is for a velocity resolution of 0.2 km.$s^{-1}$ in Stokes $I$. }         
\label{source_list}      
\centering                          
\begin{tabular}{lccccccccc}        
\hline                 
Object & Type &RA & Dec$^{a}$  & d &  ${V_{LSR}}$ & L$_{\textrm{bol}}$ &  \.M &  T$_{sys}$ & rms  \\   
                      && [h m s] & [$^{\circ}$ ' ''] &   [pc] & [km s$^{-1}$]    &  [$10^3~L_{\odot}$] &  [$10^{-6}$M$_{\odot}/$yr]  & [K]  & [mK] \\
                      \hline
 IRC+10216  & AGB & 09 47 57.38 & +13 16 43.7  & 120$^{d}$ & -25.5  & 9.8$^{d}$   & 10-40$^{e}$   &  200-220 & 12-14  \\
  RW LMi  &  AGB & 10 16 02.35 &  +30 34 19.0 &    400$^{d}$  &  -1.6 &  10$^{d}$   &    5.9$^{f}$  &  200-260 &  12    \\
  RY Dra  & AGB &  12 56 25.91 &   +65 59 39.8  &   431$^{d}$ &    -7.3 &  4.5$^{d}$ &  0.2$^{g}$ &  290-360 &  60   \\ 
AFGL618  & PPN &   04 42 53.67 &    +36 06 53.2  &    900$^{a}$   &    -25.0 &    10-14$^{b}$   &    20-110$^{c}$   &    200 &    14  \\ 
  NGC7027  & PN &  21 07 01.59 &   +42 14 10.2  & 980$^{h}$ &    25.0 &    7.2$^{h}$ &   N$/$A &  230-250 &  14   \\  
\hline                                  
\end{tabular}
\tablefoot{$^{(a)}$Sanchez Contreras \& Sahai (2004). $^{(b)}$ Knapp et. al (1993). $^{(c)}$ Lee et al. (2013a,b). $^{(d)}$ Ramstedt \& Olofsson (2014). $^{(e)}$ De Beck et al. (2012). $^{(f)}$ De Beck et al. (2010). $^{(g)}$ Cox et al. (2012). $^{(h)}$ Zijlstra et al. (2008).} 
\end{small}
\end{table*}

Moreover, during this stellar evolution, the star's geometry changes drastically: the quasi-spherical object inherited from the main sequence becomes axisymmetrical, point-like symmetrical, or even shows higher order symmetries when reaching the  protoplanetary nebulae (PPN) and PN stage \citep[e.g.][]{balick2002}. The classical or Generalized Interacting Stellar Winds (GISW) models try to explain this shaping by the interaction between a slow AGB wind with a faster post-AGB wind. However, modeling of complex structures with peculiar jets is difficult. Moreover, this mechanism will amplify an initial asymmetry in the slow wind that has to be explained first. 
Understanding the launching of post-AGB jets is fundamental to understand the post-AGB evolution. In several cases, the presence of a nearby companion might produce and maintain disks and jets, an envelope rotation, and could explain how the shaping is launched \citep[e.g.][] {akashi2015, boffin2012}. There are indications of stable, probably rotating disks in some post-AGB nebulae, mainly around binary stars \citep[][]{alcolea2007, bujarrabal2013}. 

Stellar magnetism can be one of the main ingredients in the shaping process. As a catalyst and$/$or as a collimating agent it could be the cause of a higher mass-loss rate in the equatorial plane, and thus could determine the global shaping of these objects (Blackman 2009). In addition, the ejection of massive winds by AGB stars could be triggered by magnetic activity in the degenerated core as demonstrated with magneto-hydro-dynamics (MHD) simulations \citep[][]{pascoli2008, pascoli2010}.

Actually, there are several indications of the presence of a magnetic field at the stellar surface and in the CSE of these post-main sequence objects. In the Red Giant Branch (RGB) phase, magnetic fields have been detected. \citet{auriere2015} obtained 29 Zeeman detections with Narval$/$ESPADOnS in a sample of active single G-K giants revealing, for the majority of them, a dynamo-type magnetic field. \citet{konstantinova2014} have reported the detection of magnetic fields at and above the Gauss level in approximately 50\% of their RGB$/$AGB sample demonstrating that the magnetic field is commonly detected at the surface of these objects. 
Moreover, several studies have revealed the presence of a strong magnetic field in the CSE of AGB, post-AGB stars, and PNe. For O-rich AGB objects, the magnetic field strength is estimated from the polarized maser emission of several molecules, located at different distances from the central star: B $\sim$ 0-18 G with mean value of 3.5 G in the inner part at 5-10 AU (SiO masers, Herpin et al. 2006), a few 100 mG at 100 AU \citep[water masers,][]{vlemmings2001,leal2013}, and around 10 mG in the outer part at 1000-10 000 AU \citep[OH masers,][]{kemball1997,rudnitski2010}. Recently, for the first time the magnetic field strength has been estimated to be about 2-3 G at the surface of the S-type Mira star $\chi$ Cyg \citep[][]{lebre2014}. Moreover, direct observational evidence of large-scale magnetic fields at the surface and in the environment of Post-AGB stars has been recently established \citep[][]{sabin2007,sabin2014a,sabin2015}. The post-AGB study (from OH masers) of \citet{gonidakis2014} indicates that the B strength detected in CSEs increases with the post-AGB age. While no strong magnetic field has been detected in the central stars of PNe \citep[][]{jordan2012,leone2014,steffen2014}, large-scale fields have been observed in their nebulae \citep[][]{gomez2009} and are shown to be responsible for the observed small-scale structures.

Indications for a link between morphological structures and the magnetic field exist. \citet{lebre2014} have underlined a connection between the surface magnetic field and the atmospheric shock waves in $\chi$ Cyg, a star exhibiting a departure from spherical symmetry at the photospheric level \citep[][]{ragland2006}. The MHD-dust-driven modeling of \citet{thirumalai2012} shows that the magnetic field certainly plays a role in the mass loss process in the equatorial plane of the Mira star o Ceti. In the OH$/$IR star OH231.8+4.2, a binary system, the magnetic vectors follow the molecular outflows \citep[][]{sabin2014a}, the field strength being estimated to be 2.5 Gauss at the stellar surface \citep[][]{leal2012}. The same authors have shown a clear correlation between field orientation (toroidal field oriented along the equatorial torus) and the nebular structure of the PN NGC7027 \citep[][]{sabin2007}.

As for AGB stars, the origin of the field detected in the CSEs remains unclear, as well as its possible variability with time. All of the measurements performed so far throughout AGB CSEs \citep[see][]{vlemmings2012} favor a 1/r law for the radial dependence of a toroidal magnetic field. \citet{pascoli1997} and \citet{pascoli2008,pascoli2010} proposed that a toroidal magnetic field of $\sim10^6$ G is produced by a dynamo mechanism in the degenerate core and results in a field strength of a few 10 G on the stellar surface. On the opposite, the polarization morphology of SiO masers in the circumstellar envelope of an AGB star has been investigated by \citet{assaf2013} using the VLBA and appears to be consistent with a radial magnetic field. 

Another condition for the field to be able to shape stars is that it must be sustained over the AGB lifetime. This implies either the presence of a companion to spin up the envelope of the star, thus providing the missing angular momentum \citep[][]{nordhaus2007,blackman2009}, or that the differential rotation between the core and the envelope of the star has to be re-supplied via convection or another mechanism. 

In addition, some uncertainties remain concerning the magnetic field obtained throughout the CSEs of evolved stars. Indeed, the magnetic field strength (B) derived from SiO masers are still debatable because anisotropic pumping can produce strong polarized maser emission \citep[][]{western1983,desmurs2000} and magnetic field strengths of a few 15 mG might be sufficient to explain the observational results \citep[][]{houde2014}. Hence, using field tracers other than SiO is crucial in order to have a reliable estimate of B and of its variation throughout the envelope. Moreover, most past studies focused on O-rich stars and AGB objects and similar studies should be conducted for C-rich objects. However, the main probe close to the stellar atmosphere, SiO maser emission, is only present in O-rich evolved objects and disappears soon after the star has reached the end of the AGB phase \citep[][]{nyman1998}. As a consequence, other magnetic field probes, such as the CN radical used in this work, are most useful to study more advanced stages of the stellar evolution or C-rich objects.

In this paper we present CN Zeeman observations of several C-rich objects at different evolutionary stages. We first study the distribution of the magnetic field in IRC+10216 and then present the result for four other C-rich evolved stars to compare. Sections \ref{sec:sample}  and \ref{sec:observations} present our source sample and observations, respectively. Details on CN Zeeman splitting and on data analysis are given in Sect. \ref{sec:data_analysis}. Results from the Zeeman interpretation are given in Sect. \ref{sec:results}. We finally discuss the importance of the magnetic field and compare our results with previous studies in Sect.\ref{sec:discussion}.

%

\section{Source sample}
\label{sec:sample}

\begin{table}
\centering
\caption{For each object in our sample we give (from the literature, see Sect. \ref{sec:sample}) for the CN layer, the molecular abundance relative to  H$_2$, its distance d$_{CN}$ to the central star, the stellar radius.}
\begin{tabular}{lccc}
\hline
 Object & $[$CN$/$H$_2]$ & d$_{CN}$ & R$_{*}$   \\
 & & [AU] & [AU]  \\
\hline
RW LMi &  $3.$ $10^{-5}$ & 2675-3340 (3-9'') & 2.6  \\
RY Dra & $5.1$ $10^{-5}$ & 61-615 (0.14-1.5)" & 1.0 \\
IRC+10216  & $8.0$ $10^{-6}$ & 2500 (21'') & 3.3  \\
AFGL618 & $2.1$ 1$0^{-6}$ & 2700 (3'') & 0.24  \\
NGC7027 & $2.3$ $10^{-7}$ & 10000 (11'') & 3.5 $\times 10^{-4}$   \\
\hline
\end{tabular}
\label{CN}
\end{table}

Our source sample consists of five evolved low- or intermediate-mass carbon stars: three AGB objects (IRC+10216, RY Dra and RW LMi), one proto-PN (AFGL618), and one PN (NGC7027). The stellar parameters (coordinates, distance, LSR velocity, bolometric luminosity, estimated mass-loss rate) are given in Table \ref{source_list} and each source is presented below. Most of these sources have already been observed in CN by several authors; we try to estimate the size  of the CN layer and the distance to the central object for each star (see Table \ref{CN}). 

IRC+10216 is one of the closest AGB stars \citep[120 pc, period of 630 days,][]{ramstedt2014} and the best studied C-rich object. Half of the known interstellar species are observed in its outer envelope. It is surrounded by an optically thick C-rich CSE, which results from the ejection of stellar material at a rate of $1-4 \times 10^{-5}$ M$_{\odot}/$yr \citep[][]{debeck2012}. At first glance, IRC+10216 is nearly spherical and expands radially (over more than 450$\arcsec$ in CO) with a velocity of 14.5 km.s$^{-1}$ \citep[][]{cernicharo2015}. Deviations from symmetry are visible at small angular scales \citep[e.g.][]{skinner1998}, suggesting the presence of an overall bipolar structure. Moreover, high angular resolution observations indicate the presence of clumps and show that the innermost structures at subarsecond scale are changing on timescales of years \citep[e.g.][]{tuthill2000}, and that this object might have begun its evolution towards the PN phase.  ALMA observations by \citet{decin2015} have revealed that the bipolar structure with concentric shells is indeed a binary-induced spiral shell. A faint companion of the AGB star might have been discovered by \citet{kim2015} and may explain the observed circumstellar geometry. However, a cyclic magnetic activity at the stellar surface, similar to that on the Sun, is an alternative explanation \citep[][]{soker2000}. Investigating the magnetic field strength in IRC+10216 might help to discriminate between these two alternatives. A large 40\arcsec diameter ($\sim$ 5000 AU) CN ring with two symmetrical brighter emission lobes has been mapped with the IRAM interferometer \citep{lucas1995}. In addition, \citet{lindqvist2000} estimated the CN abundance to $8\times 10^{-6}$ and inferred a CN ring with a radius of $5\times 10^{16}$ cm ($\sim$3300 AU) and a width of $4\times 10^{16}$ cm. 

Two other C-rich AGB objects, RW LMi and RY Dra, are studied in addition to IRC+10216. RW LMi, also known as CIT6, is an M-type C-rich star \citep[period of 640 days,][]{ramstedt2014} with a very rich circumstellar envelope \citep[][]{schmidt2002}, believed to be in transition from the AGB to the post-AGB phase. \citet{schmidt2002} found the presence of a nascent bipolar nebula, providing evidence that the evolutionary phase of CIT6 lies just past the tip of AGB. Like IRC+10216, some spiral structure has been discovered in the circumstellar envelope. It is likely induced by a central binary star \citep[][]{kim2013}, and it extends over 20$\arcsec$. Actually, the molecular content of the RW LMi envelope suggests that this object is more evolved than IRC+10216 \citep[][]{chau2012}. The interferometric observations of \citet{lindqvist2000} reveal a CN ring with a radius of 2675-3340 AU. 
RY Dra belongs to the J-type carbon stars (i.e. stars with $^{12}$C$/^{13}$C-ratios $\sim3$). This object is a b-type semi-variable \citep[period= 173 days,][]{ramstedt2014}. A detached shell has been revealed by the ISO observations of \citet{izumiura1999} and could have been produced by a previous episode of mass loss. Neither interferometric data nor Hubble images are available for this object. The PACS Herschel map (Herschel archive) of this object does not spatially resolve the spherical circumstellar envelope whose size appears to be a few arcsec. Only CN single dish observations are available for RY Dra \citep[][]{bachiller1997b} which lead to a CN abundance of $5.1\times 10^{-5}$ and an estimated CN ring size of 0.14-1.5\arcsec~(61-615 AU). 
 
AFGL618, also named the Westbrook Nebula, is a young proto-PN \citep[$\sim$200 years][]{kwok1984} exhibiting a B0 central star, a central compact HII region \citep[][]{sanchez2002}, and two pairs of rapidly expanding well collimated lobes \citep[][]{balick2013}. SMA continuum and CO line polarization observations of \citet{sabin2014a} have revealed a magnetic field well aligned and organized along the polar direction, suggesting a magnetic outflow launching mechanism. The extended ($5\arcsec$) molecular envelope is composed of material ejected during the AGB phase and partly chemically reprocessed \citep[e.g.,][]{herpin2002}. The molecular outflows extend over 20$\arcsec$ \citep[][]{sanchez2002}. Two episodes of collimated fast winds have been identified \citep[][]{lee2013b}: one with a huge mass-loss rate of $\sim1.1~\times~10^{-4}$ M$_\odot$yr$^{-1}$, and an older one with $\sim2~\times~10^{-5}$ M$_\odot$yr$^{-1}$. The CN abundance ($2.1\times 10^{-6}$) has been derived by \citet{bachiller1997}. To estimate the CN distribution and abundance in this object, an HCN map only is available \citep[][]{sanchez2002}. Since HCN is a molecule of photospheric origin that gets photodissociated by the ambient interstellar UV-field into CN \citep[e.g][]{huggins1982} we can assume the CN molecules are surrounding the HCN envelope, leading to a rough estimate of the CN envelope size of 6\arcsec~in diameter, hence $8.1\times10^{16}$cm (or 5400 AU).

NGC7027 is a young PN \citep[kinematical age of 600 years,][]{masson1989}, with a high degree of axial symmetry. Recent studies have shown that this object is a multipolar PN in the making \citep[see][]{huang2010}. High-velocity jets exhibit a multipolar shape in H$_2$ extending over roughly 20$\arcsec$ \citep[][]{sabin2007}. A spherical CO envelope extends over 60 $\arcsec$ around the central HII region. While the inner part of the torus has been ionized, the ancient AGB molecular content has been completely reprocessed in the envelope \citep[][]{herpin2002}. The HCN emission extends over 20$\arcsec$ \citep[][]{huang2010}. We had access to an estimate of the CN distribution from Plateau de Bure observations (Josselin et al., private communication,), showing CN at a a distance of 10000 AU from the central object.

%

\section{Observations}
\label{sec:observations}

We have obtained simultaneous spectroscopic measurements of the 4 Stokes parameters \textit{I, U, Q, V} for the seven CN 1-0 hyperfine transitions given in Table \ref{zeeman}. 
Two observing runs have been performed, the first one in November 2006 and the second one in March and June 2016. The observations have been carried out with the XPol polarimeter \citep[][]{thum2008} at the IRAM-30m telescope on Pico Veleta, Spain. 

For the first observing session (November 2006), the pointing was regularly checked directly on the observed stars themselves. The system temperature of the SIS receiver and the rms are given in Table \ref{source_list}. The front-ends were the facility receivers A100 and B100, and the back-end was the VESPA correlator. The lines were observed with a spectral resolution of 0.11 km.s$^{-1}$ (40 kHz) in order to observe the Zeeman effect with a sufficient accuracy. All hyperfine components lines were simultaneously covered with two VESPA sections of 80 and 40 MHz width, respectively, each one being correctly centered. The integration times (pointed, wobbler-switched observations) were 137, 175, 248, 17, and 215 minutes, for AFGL618, IRC+10216, RW LMi, RY Dra, and NGC7027 respectively. The forward and main beam efficiencies were 0.95 and 0.74, respectively, at the CN frequencies, while the half-power beam width was 21.7 \arcsec. The Jy$/$K conversion factor is 6.3. The polarization angle calibration \citep[see][]{thum2008} has been verified by means of observations of the Crab Nebula. Moreover, planets have been used to check the instrumental polarization along the optical axis (their intrinsic polarization is negligible at the considered frequency, or, for Mars and Mercury, cancels out in the beam). A more detailed discussion about any instrumental effect is given in Sect. \ref{instr_cont}.

The 2016 observations were exclusively dedicated to IRC+10216. The main goal was to map the magnetic field in the CN envelope which is resolved by the beam of the telescope. In addition to the positions observed in 2006 ((0\arcsec, 0\arcsec), corresponding to the central source position, and (-10\arcsec, -20\arcsec), (+15\arcsec, -15\arcsec)) we observed four offset positions (-18\arcsec, -10\arcsec), (-18\arcsec, +10\arcsec), (+18\arcsec, -04\arcsec), and (+20\arcsec, +16\arcsec).
The EMIR band E090, with the backend VESPA, was set up to observe all CN 1-0 hyperfine components simultaneously, as with the facility receiver in 2006, but now with larger bandwidths (160 and 80 MHz) in order to better define the spectral baseline. The focus and the pointing were checked on Jupiter. A major change occurred at the telescope during the winter before our 2016 observations. The EMIR 3mm band was upgrated in November 2015. The mixer was exchanged, but also the dual-horn system was changed to a single horn system. The linear horizontal and vertical polarization splitting are not obtained from a grid anymore but from a orthomode transducer. The receiver has then one single horn followed by a wave guide wherein the signal is separated into horizontal and vertical components. In Sect. \ref{sec:specific} we discuss the impact of this new design on our 2016 observations, i.e. a leakage of the Stokes \textit{I} into the \textit{V} signal. One of the consequences is that we had to increase the integration time per point (on average 11 hours by position compared to 1 hour before) to retrieve a sufficient S/N in the \textit{V} signal.

\begin{table}
\centering
\caption{Zeeman splitting factor $Z$ for CN N=1$\rightarrow$0 hyperfine components (Crutcher et al. 1996). R.I. stands for Relative Intensity in LTE conditions. Two CN groups are distinguished: lines 1-3 and 4-7}
\begin{tabular}{p{0.01cm}cccp{0.3cm}c}
\hline
\#& $N'_{J',F'}\rightarrow N_{J,F}$ & $\nu_{0}$ & $Z$  & R.I. & $Z\times R.I.$ \\
 & & [GHz] & [Hz$/\mu G$] & & \\
\hline
1 & $1_{1/2,1/2}\rightarrow 0_{1/2,3/2}$ & 113.14434 & 2.18 & 8 & 17.4 \\
2 & $1_{1/2,3/2}\rightarrow 0_{1/2,1/2}$ & 113.17087 & -0.31 & 8 & - 2.5 \\
3 & $1_{1/2,3/2}\rightarrow  0_{1/2,3/2}$ & 113.19133 & 0.62 & 10 & 6.2 \\
\hline
4 &  $1_{3/2,3/2}\rightarrow 0_{1/2,1/2}$ & 113.48839 & 2.18 & 10 & 21.8 \\
5 & $1_{3/2,5/2}\rightarrow 0_{1/2,3/2}$ & 113.49115 & 0.56 & 27 & 15.1 \\
6 & $1_{3/2,1/2}\rightarrow 0_{1/2,1/2}$ & 113.49972 & 0.62 & 8 & 5.0 \\
7 &  $1_{3/2,3/2}\rightarrow 0_{1/2,3/2}$ & 113.50906 & 1.62 & 8 & 13.0 \\
\hline
\end{tabular}
\label{zeeman}
\end{table}

%

\section{Data analysis and method}
\label{sec:data_analysis}
\subsection{CN and magnetic field}
\label{sec:CN}

   \begin{figure*}
   \includegraphics[width=525pt]{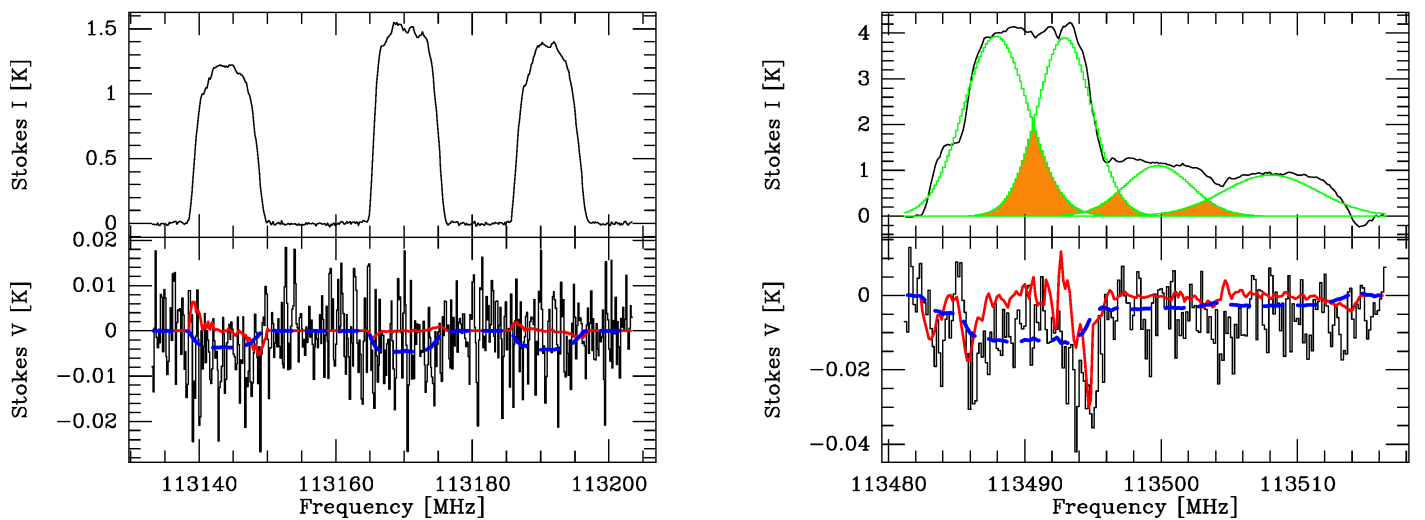}
   \caption{ IRC+10216: observations of November 2006 for the position $(-10\arcsec, +20\arcsec)$. {\em Top Left}: CN (1, 1/2) $\rightarrow$ (0, 1/2) Stokes \textit{I} spectrum. {\em Bottom Left:} \textit{V} spectrum (in black) and least-squares fit (in red) using the \citet{crutcher1996} method described in text. The dashed blue line shows our fit assuming C$_3$ = 0. \textit{Right}: same for the CN transition (1, 3/2) $\rightarrow$ (0,1/2), with in addition the Gaussian fit for each individual line overplotted in green and the line overlap area colored in orange on the \textit{I} spectrum.}
              \label{magnetic field IRC-10+20i}%
    \end{figure*}

Observing CN offers a good opportunity to measure magnetic fields in carbon stars. First of all, because of the large abundance of the CN radical \citep[up to $\sim10^{-5}$,][]{bachiller1997b, bachiller1997}, the N=1$\rightarrow$0 and N=2$\rightarrow$1 lines have already been observed and easily detected in several carbon-rich AGB stars and PNe \citep[e.g.,][]{bachiller1997b, bachiller1997,josselin2003}.
Moreover, CN is a paramagnetic species, thus exhibiting Zeeman splitting when the spectral line-forming region is permeated by a magnetic field B. The only currently viable technique for measuring the magnetic field strength in the circumstellar envelope of carbon stars is to detect the Zeeman effect in spectral lines excited in the envelope. The normal Zeeman effect splits a line with rest frequency $\nu_{0} $ into three separated polarized components with frequencies $\nu_{0} - \nu_{z}$, $\nu_{0}$ and $\nu_{0} + \nu{z}$, where $2\nu_{z} = \mid B \mid Z$, where $Z$ is the Zeeman factor.

The CN N=1-0 line has a total of nine hyperfine components split into two groups (one around 113.17 GHz and the second one around 113.49 GHz), with seven main lines, out of which four exhibit a strong Zeeman effect (see last column of Table \ref{zeeman}, lines 1, 4, 5, and 7).

\subsection{Analysis method}
\label{sec:methods}

\subsubsection{Data reduction}

An electromagnetic wave is defined by its horizontal and vertical components:
\begin{equation}
\label{ }
e_H(z,t)=E_H\ e^{j(\omega t-kz-\delta)} 
\end{equation}
\begin{equation}
\label{ }
e_V(z,t)=E_V\ e^{j(\omega t-kz)}
\end{equation}
where $\delta$ is the phase difference between horizontal and vertical components.

For each source observed in polarimetry at the 30m and each VESPA section, i.e. each CN hyperfine lines group (lines 1-3 and 4-7 in Table \ref{zeeman}), the spectrometer output is converted to the Stokes parameters as defined in the equatorial reference frame (i.e., counting the polarisation angle from North to East):
\begin{equation}
\label{ }
I = <{E_H}^2> + <{E_V}^2>
\end{equation}
\begin{equation}
\label{ }
Q = <{E_H}^2> - <{E_V}^2>
\end{equation}
\begin{equation}
\label{ }
U = 2 <E_H E_V cos\  \delta>
\end{equation}
\begin{equation}
\label{ }
V = 2 <E_H E_V sin\  \delta>
\end{equation}

All data are reduced using the CLASS software\footnote{http://www.iram.fr/IRAMFR/GILDAS/}. All \textit{I}, \textit{Q}, \textit{U}, and \textit{V} spectra have been inspected individually. A few of them have been discarded due to technical problems. A baseline has been removed from all \textit{I, U, Q}, and \textit{V} spectra (excluding in the \textit{U} and \textit{V} spectra the frequency range where the CN line emission in Stokes $I$ is above the noise) using an order two polynomial.

\subsubsection{Numerical method for \textit{V} spectra}
\label{sec:crutcher}

In this subsection we use the Stokes \textit{I} and \textit{V} spectra obtained after data reduction, as shown in Fig. \ref{magnetic field IRC-10+20i}. To determine the magnetic field, \citet{crutcher1996} developed a procedure to fit all seven hyperfine components in the Stokes $V$ spectra in the least-squares sense: 
\begin{equation}
V_{i}(\nu) = C_{1}I_{i}(\nu) + C_{2}\frac{dI_{i}(\nu)}{d\nu} + C_{3}Z_{i}\frac{dI_{i}(\nu)}{d\nu}
\label{crutchereq}
\end{equation}
with i = 1 to 7 for the seven hyperfine components exhibiting a strong Zeeman effect.

In this expression \textit{$V_{i}(\nu)$} and \textit{$I_{i}(\nu)$} are the Stokes \textit{V} and \textit{I} spectra for each of the seven hyperfine CN lines.
This method accounts for the Zeeman and instrumental effects. The instrumental contribution is determined by $C_{1}$ and $C_{2}$. $C_{1}$  depends on the gain difference in the telescope between the right (R) and left (L) circular polarizations. $C_{2}$ accounts for apparent Zeeman effect due to telescope beam squint when it observes sources with a velocity gradient, or to residual instrumental effects leading to residual frequency offset between R and L. 
Hence, we can estimate the line of sight component of the magnetic field (B$_{los}$) from $C_{3} = \frac{B_{los}}{2}$, once $C_{1}$ and $C_{2}$ have been calibrated.

We have developed a fitting procedure, applied simultaneously to the seven hyperfine components, which slightly differs from the "Crutcher technique" in that we do not fit for the three $C_{i}$ parameters simultaneously but separately first. A first guessed value is attributed to the parameters $C_1$ and $C_2$. In an iterative process, we then explore a large parameter space simultaneously for both $C_1$ and $C_2$ in order to get a rough estimate of the instrumental contribution. Next, still allowing $C_1$ and $C_2$ to vary (but within a smaller range of values), the parameter space for C$_3$ is investigated (in an iterative process too) while \textit{V} is simultaneously calculated for the seven transitons. Convergence of the iterative process is checked by using a $\chi^2$ method and is reached when $ 0.18 < \chi^2 < 0.23$ for our 2006 data and $ 0.11  < \chi^2 < 0.17$ for our 2016 data.

The main issue of this method is to determine the frequency range corresponding to the emission of each CN hyperfine component. While the components of the first group (lines 1-3 in Table \ref{zeeman}) are spectrally well identified (Fig.\ref{magnetic field IRC-10+20i} left part), lines 4-7 are blended for the sources studied here (Fig.\ref{magnetic field IRC-10+20i} right part). It is thus necessary to determine the frequency intervals where the overlap occurs. For these intervals a blend of individual Stokes features is seen, i.e., $V_4$ and $V_5$ for the CN hyperfine components 4 and 5, respectively: 
\begin{equation}
V_4+V_5=C_3 (Z_4 \frac{dI_{i}(\nu)}{d\nu} + Z_5 \frac{dI_{i}(\nu)}{d\nu})
\end{equation}
 These overlapping frequencies (plotted in orange in Fig.\ref{magnetic field IRC-10+20i}, right part) are derived from a Gaussian fit (using CLASS) of the CN (1-0) lines profiles (see Fig.\ref{magnetic field IRC-10+20i} right part, in green).

\subsubsection{Specific treatment for the 2016 observations}
\label{sec:specific}

As explained in Sect. \ref{sec:observations}, in November 2015, the 3mm mixer and the 30m optics were modified. The dual-horn system was changed to a single horn system and the vertical (V) and horizontal (H) polarization splitting is produced now by a orthomode transducer. Using a single horn instead of two different horns for H and V should eliminate the misalignment of the H and V horns as a major contribution to the instrumental polarization. 

\begin{figure}
   \includegraphics[width=190pt, angle=270]{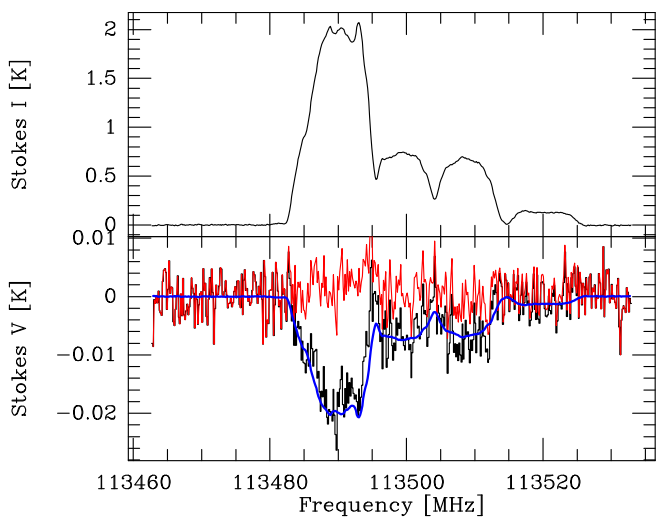}
   \caption{IRC+10216: observations of March and June 2016 for position (+20\arcsec, +16\arcsec). CN (1, 3/2) $\rightarrow$ (0, 1/2) Stokes \textit{I} (\textit{Top}) and \textit{V} (\textit{Bottom}) spectra in black and fit of the estimation of the \textit{I} leakage in blue. The resulting true \textit{V} spectrum after removal of the \textit{I} leakage is shown in red.}
              \label{leaki}
    \end{figure}

According to the IRAM technical tests (IRAM internal communication), the instrumental contamination in the linear polarization is improved and the beam squint effect has been removed. However, our March 2016 observations showed that the \textit{V} instrumental polarization has strongly increased. This change has produced a substantial leakage of the Stokes \textit{I} signal into the Stokes \textit{V}, as revealed by the mirror image of \textit{I} seen in the \textit{V} spectra (see Fig. \ref{leaki}).

We now assume about the relative importance of the three $C_{i}$ parameters in formula (\ref{crutchereq}). As the beam squint effect is now cancelled, we assume that the $C_2$ coefficient is equal to zero and that we are dominated by the leakage of the $I$ into $V$, i.e. $C_1$ . To estimate the leakage of \textit{I} into \textit{V}, measured by the $C_1$ coefficient, we first assume that the magnetic field is negligible with respect to the leakage, i.e. $C_3 = 0$. Hence, formula (\ref{crutchereq}) becomes:
\begin{equation}
 V=V_{cal_{i}} (\nu) = C_{1}I_{i}(\nu)
 \end{equation}
 
 $C_1$ gives the percentage of the \textit{I} into \textit{V} leakage for the observations of IRC+10216 performed in March and June 2016. $C_1$ is found around 1.1 and 1.4 $\%$ (see Table. \ref{result_IRC}) depending on the observed positions. The resulting $V_{cal}$ spectra is shown in blue for one observed position in Fig. \ref{leaki}. No correlation has been found between $C_1$ and $I$ or the degree of elevation of the source. 
 
Finally, to estimate the true \textit{V} signal (\textit{V$_{true}$}, i.e. not contaminated by \textit{I}) we then subtract \textit{V$_{cal}$} from the original \textit{V$_{ori}$} spectra, that is to say: 
\begin{equation}
V_{true}=V_{ori} -V_{cal_{i}} (\nu)=V_{i} -C_{1}I_{i}(\nu)
\end{equation}
This leakage-free $V$ spectrum, $V_{true}$, is plotted in red in Fig. \ref{leaki}. Then, we can apply the analysis described in Sect.\ref{sec:crutcher} on this \textit{V$_{true}$}, signal, to derive the C$_3$ coefficient. 

   \begin{figure}
   \includegraphics[scale=0.35]{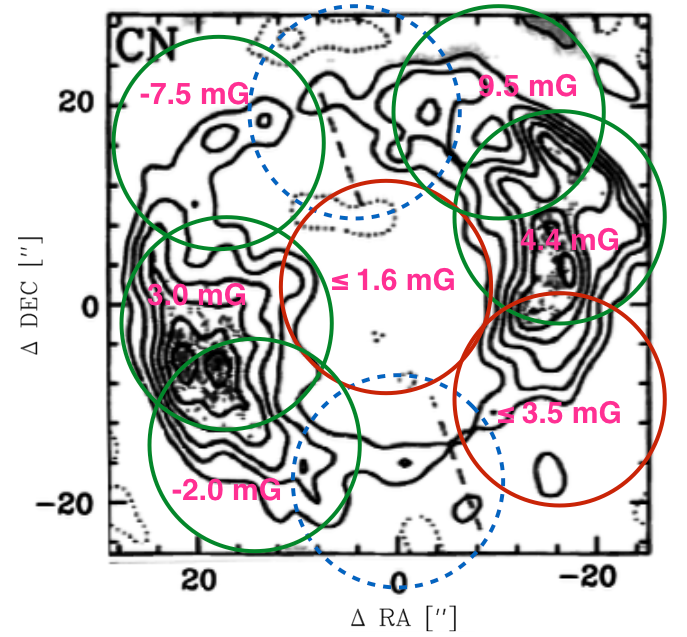}
   \caption{CN map of IRC+10216 adapted from \citet{lucas1995}. Green, red and dashed blue circles represent positions with a Zeeman detection, without a Zeeman detection, and unobserved ones, respectively. The diameter of the circles corresponds to the size of the telescope primary beam $(21.7\arcsec)$.}
              \label{mapIRC}%
    \end{figure}

\section{Results}
\label{sec:results}

\subsection{Uncertainties and method limitations}

\label{instr_cont}
As a consequence of formula (7), the measured accuracy of the magnetic field (B$_{los}$) strength depends on 1) the rms of the \textit{I} and \textit{V} observations and 2) the precise characterization of the instrumental contamination. 

\subsubsection{Zeeman features in the Stokes V spectra}
\label{V_rms}

\begin{figure}
   \centering
   \includegraphics[width=220pt, angle=270]{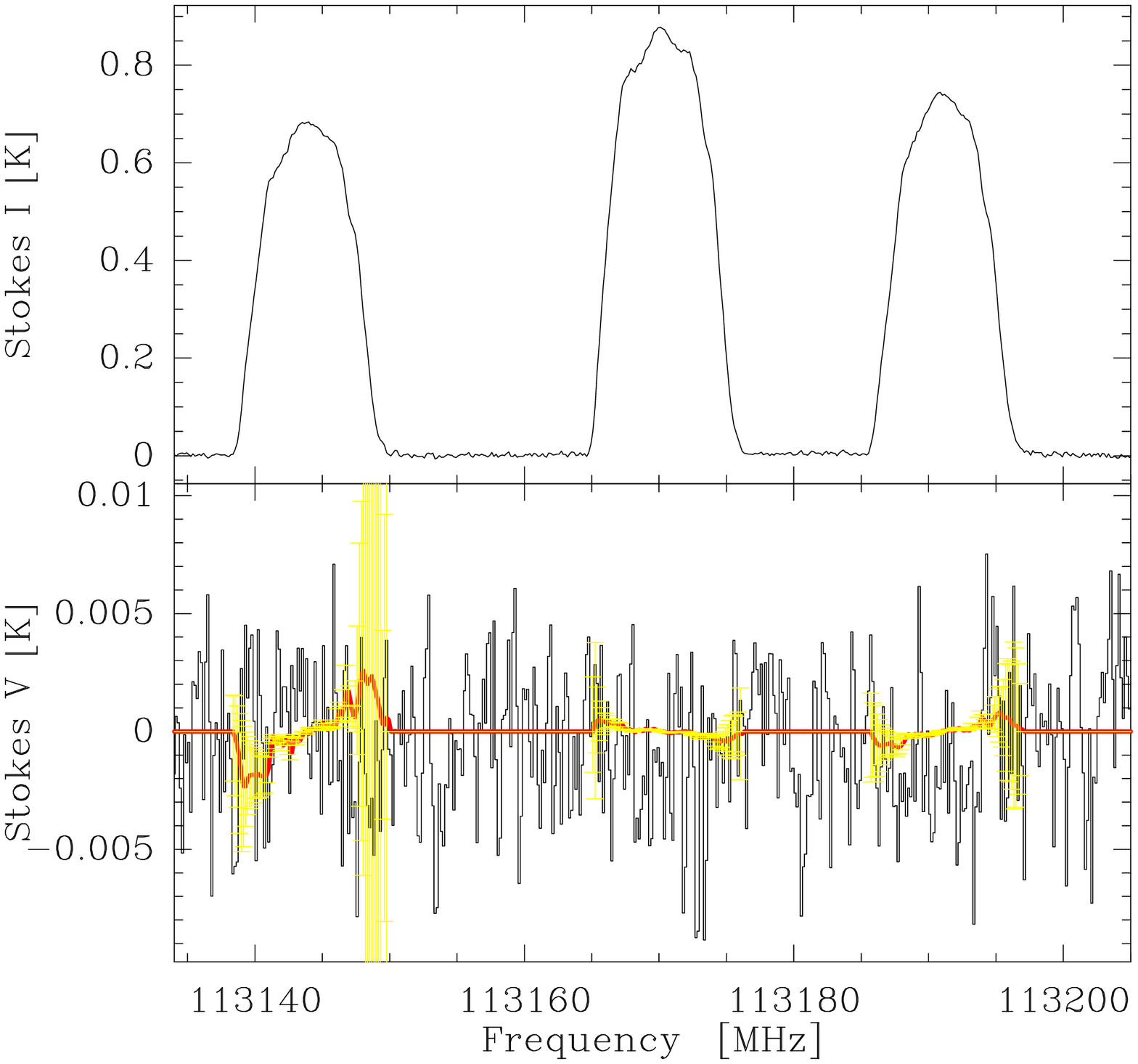}
   \includegraphics[width=220pt, angle=270]{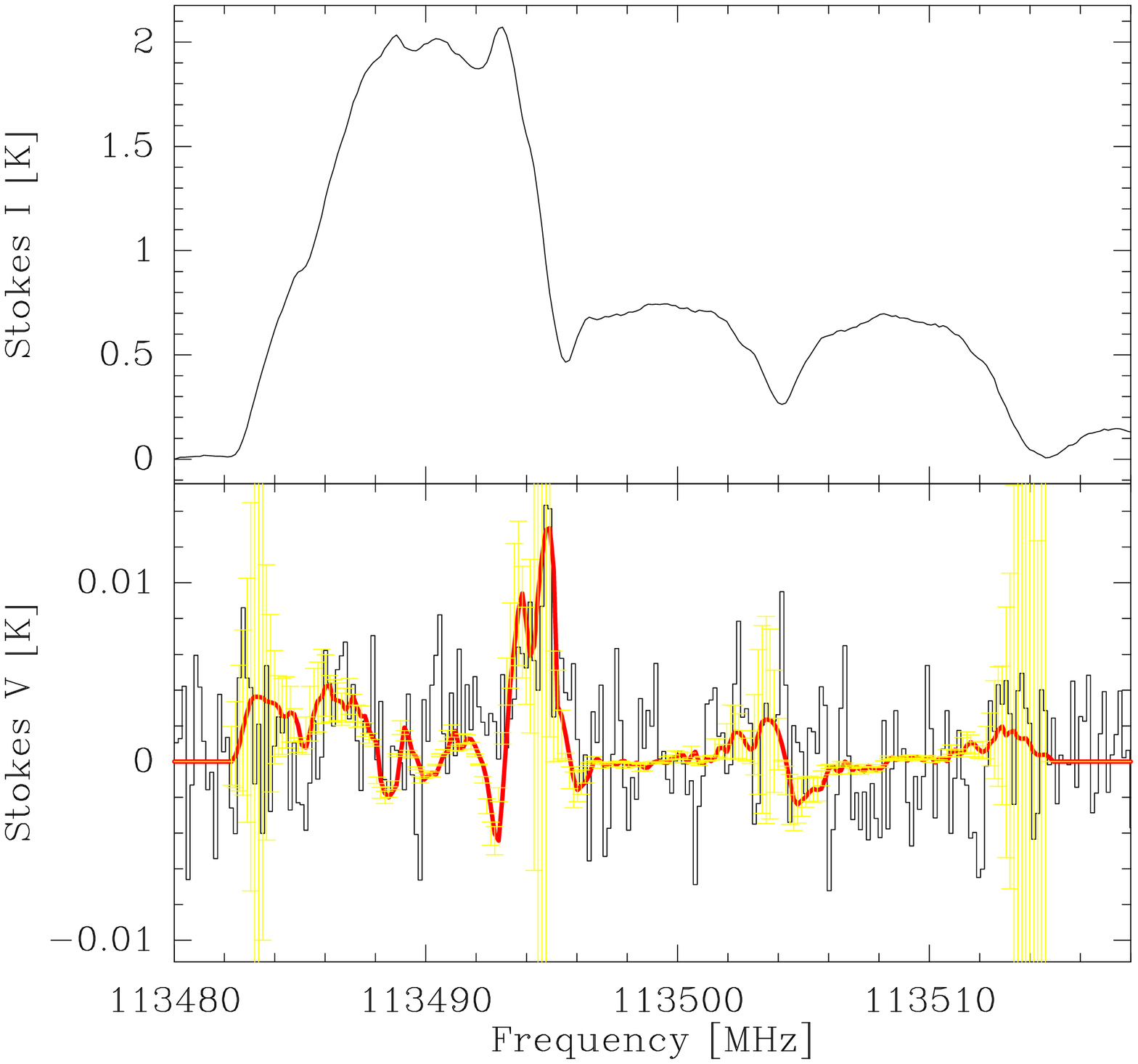}
   \caption{ IRC+10216: observations of March and June 2016 for the position (+20\arcsec, +16\arcsec). \textit{Top:} Stokes \textit{I} and \textit{V} for CN transition (1, 1/2) $\rightarrow$  (0, 1/2). \textit{Bottom:} same for CN transition (1, 3/2) $\rightarrow$ (0,1/2). Spectra and least-squares fits for \textit{V} are shown in black and red, respectively. The error for the Stokes \textit{V} fit is plotted in yellow.}
              \label{error2016V01}%
    \end{figure}

For all sources$/$positions the rms (for a velocity resolution of 0.2 km$/$s) of the \textit{I} and \textit{V} observations is in the range 9-14 mK, except for RY Dra for which values are 4.5 times larger. The resulting S$/$N (computed from the V integrated area divided by rms$\times \delta$v) for the likely Zeeman fetaures in the Stokes \textit{V} spectra is less than 3 for RW LMi, RY Dra, and NGC7027 while it is more than 5 for AFGL618. Concerning IRC+10216, for five of the observed positions, the Zeeman effect is likely detected (considering the leakage-free $V_{true}$ spectrum for the 2016 data) with a S/N higher than 5 (see Figs. \ref{magnetic field IRC-10+20i}, \ref{leaki}, \ref{error2016V01}, \ref{error2006V01}, and \ref{magnetic field IRC+15-15}-\ref{fieldRC6}). The two positions with the highest S$/$N, $(-10\arcsec, +20\arcsec)$ and $(+20\arcsec, +16\arcsec)$ are shown in Fig. \ref{magnetic field IRC-10+20i} and \ref{error2016V01}. No signal is detected for the central position while the S$/$N for position $(-18\arcsec, -10\arcsec)$ is less than 3. Only upper limits for B will then be inferred when the S$/$N is less than 3. 

\subsubsection{Accuracy of B strength estimates}

As explained in Sect. \ref{sec:methods}, we have estimated through an iterative process the polarization instrumental contributions from the total intensity, accounting for the leakage of \textit{I} into Stokes \textit{V}, and from the beam squint effect for all observed sources$/$positions in our sample (see Tables \ref{result_IRC} and \ref{result}). The beam squint leakage must be considered when horizontal and vertical polarization signals are not collected in the same horn, i.e. do not probe exactly the same region, and for sources with a non-zero velocity gradient. This effect, quantified by the $C_2$ coefficient, is noticeable in our first set of observations but has disappeared in the second run after a single horn has been installed on the 30m telescope. Hence, while $C_2$ is zero for the March 2016 observations, we obtain large values for the other observations. The $C_1$ parameter varies between $5.5 \times 10^{-4}$ and $3.0 \times 10^{-3}$ for the first set of observations while it reaches value up to $1.4 \times 10^{-2}$ after the strong leakage of \textit{I} into \textit{V} occurred. 


Considering the rms of the V spectra, we have estimated that the accuracy on our estimate of $C_1$ and $C_2$ is roughly 10\% and 300 Hz respectively. From equation (\ref{crutchereq}), 
\begin{equation}
C_{3}Z_{i}\frac{dI_{i}(\nu)}{d\nu}= V_{i}(\nu) - C_{1}I_{i}(\nu) - C_{2}\frac{dI_{i}(\nu)}{d\nu}  
\end{equation}
and, assuming that the error on $dI_{i}(\nu)/d\nu$ is negligible compared to the others, we can estimate the error on $C_3$ from: 
\begin{equation}
\delta (C_{3}Z_{i}\frac{dI_{i}(\nu)}{d\nu})= \delta V_{i}(\nu) - \delta C_{1} I_{i}(\nu)  - C_{1} \delta I_{i}(\nu)- \delta C_{2}\frac{dI_{i}(\nu)}{d\nu}  
\end{equation}
The error on the observed $V_{i}(\nu)$ and $I_{i}(\nu)$ being the spectral rms ($rms_V$ and $rms_I$), one derives the uncertainty on $C_{3}$, hence on the magnetic field strength:
\begin{equation}
\delta C_{3}= \frac{1}{Z_{i}\frac{dI_{i}(\nu)}{d\nu}} \times (rms_V - \delta C_{1} I_{i}(\nu)  - C_{1} rms_I- \delta C_{2}\frac{dI_{i}(\nu)}{d\nu})  
\end{equation}

It is in principle possible to relate C$_1$ and C$_2$ to the Stokes \textit{V} and Stokes \textit{I} power patterns of the telescope. Thanks to the orthomode transducer, the power pattern for the instrumental conversion of Stokes \textit{I} into Stokes \textit{V} is almost axially symmetric. This justifies our assumption that C$_2$ is insignificant in our 2016 observations. A comparison between C$_1$ and the power patterns measured on Uranus can be found in Appendix \ref{appendixB}.

In addition, for the central position (0\arcsec,0\arcsec) of IRC+10216, our spectra are slightly contaminated by the CN ring emission (see Fig. \ref{mapIRC}). A fraction of the signal is probably contaminated by the secondary telescope lobes which are located on the CN envelope (see Appendix \ref{appendixB}).

Of course, a strong limitation of these measurements is that the magnetic field strength is only measured along the line-of-sight. As a consequence finding a zero magnetic field does not necessarily mean that there is no magnetic field, i.e. the sum only of the magnetic field vectors in case of a twisted field could be zero within the telescope beam. Even if interferometric mapping of the Zeeman effect would give a higher angular resolution map of the line-of-sight component and then help to constrain this possibility, it will not give us access to the full magnetic vector. To further investigate this issue interferometric observations would nevertheless be helpful, but the spectral line polarimetry (with all Stokes parameters) is still not available with ALMA or NOEMA.

\subsection{Mapping the magnetic field in IRC+10216}
\label{sec:IRC}

IRC+10216 is an AGB carbon star whose CN ring diameter is larger than the 30m beam. For that reason, IRC+10216 is the best candidate, to obtain for the first time, a map of the magnetic field in the envelope of an evolved star. Nine positions have been proposed to cover the whole CN ring with half-beam spacing (see Fig. \ref{mapIRC}). Unfortunately, due to the weather and increasing observing time because of the Xpol issue (see Sect. \ref{sec:specific}), only 7 out of the 9 positions have been observed, three during the first run in 2006, and four in 2016. Results are presented in the Table \ref{result_IRC} (the {\em CN cartography}).

\begin{table}
\centering
\caption{ {\em CN cartography of IRC+10216}: for each position, $B_{r_{*}}$ is the extrapolated strength of the magnetic field (following a $1/r$ law) at one stellar radius (the CN layer is at 2500 AU, i.e. 21'', the stellar radius being 3.3 AU).}
\begin{tabular}{lccccc}
\hline
 Position  & $B_{los}$  &  $\delta B_{los}$ & $\mid B_{r_{*}} \mid$ & $\mid C_1 \mid$ & $C_2$ \\ 
arcsec & [mG] & [mG] & [G]  & $\times 10^{-3}$ & [Hz]  \\
\hline
+0 +0$^{a}$ &  $\leq 1.6$ &  &$\leq 1.1$  & $1.05$ & 530 \\
-10 +20$^{a}$ & 9.5 & 5.5 & 7.2  & $3.0$ & 800 \\
+15 -15$^{a}$ & -2.0 & 6.7 &1.5 & $0.55$  &750 \\ \hline
-18 +10 & 4.4 & 1.8  & 3.3  & $12$ & no \\
-18 -10 & $\leq 3.5$ &   & $\leq 2.7$  & $12$ & no \\
+18 -04 & 3.0 & 0.5 & 2.2  & $14$ & no \\
+20 +16 & -7.5 & 1.2 & 5.7  & $10$ & no \\
\hline
\end{tabular}
\tablefoot{$^{(a)}$ Observations made in 2006 before the optics modification (see section \ref{sec:methods}).} 
\label{result_IRC}
\end{table}

\begin{figure}
   \centering
   \includegraphics[width=220pt, angle=270]{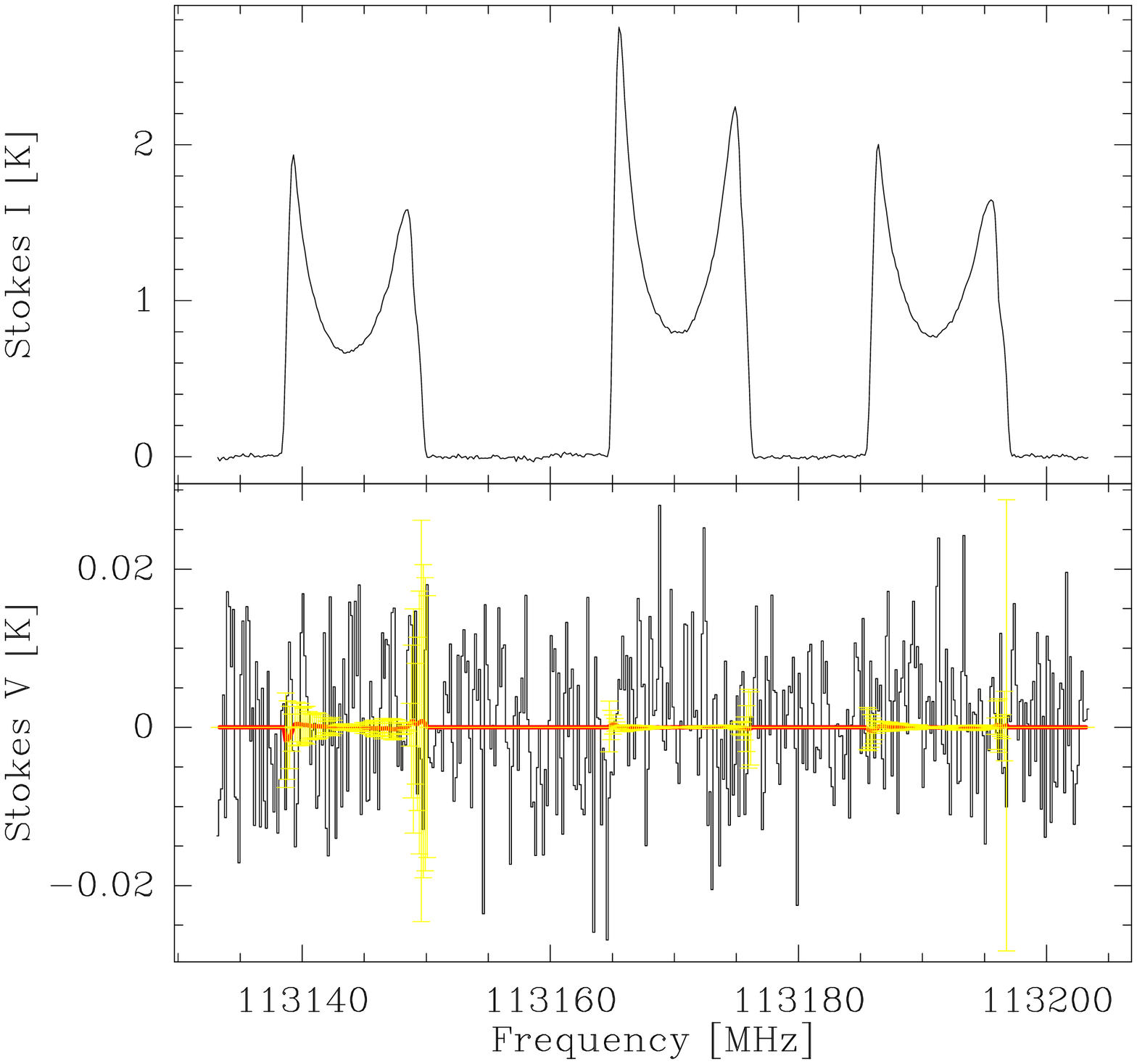}
   \includegraphics[width=220pt, angle=270]{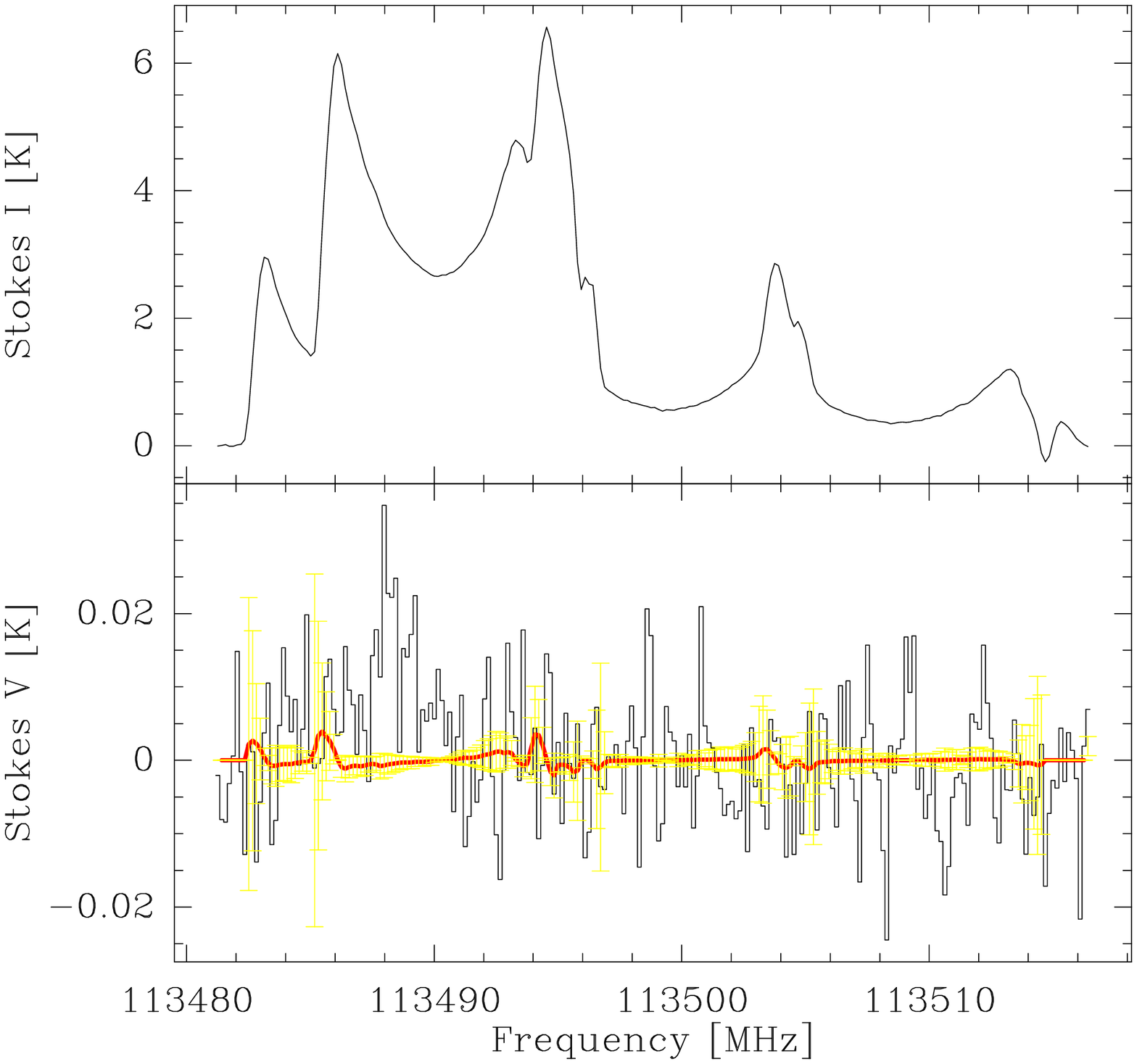}
   \caption{As in Fig. \ref{error2016V01} for observations of November 2006 for the position (0\arcsec, 0\arcsec) toward IRC+10216.}
              \label{error2006V01}%
    \end{figure}

For the central position, we first note that the Stokes \textit{I} line profiles exhibit a double horn profile for each CN component (see  Fig. \ref{error2006V01}), meaning that the CN is expanding. We measured an expansion velocity of about 14.0 km.s$^{-1}$, in agreement with \citet{fong2006}.  On this central position (see Fig. \ref{error2006V01}) and on $(-18\arcsec, -10\arcsec)$, the \textit{V} signal is dominated by the noise (see Sect. \ref{V_rms})  and only upper limits for B$_{los}$ are derived. The other positions exhibit a likely Zeeman effect with a good S$/$N (see Figs. \ref{error2016V01}-\ref{error2006V01}). Considering the instrumental contribution, we then derive a longitudinal component (absolute value) of the magnetic field between 2.0 and 9.5 mG depending on the position (see Table \ref{result_IRC}), with uncertainties varying from 15\% (position $(+20\arcsec, +16\arcsec)$) to 60\% (position $(-10\arcsec, +20\arcsec)$). The error on the measurement for positions $(+15\arcsec, -15\arcsec)$ is more than 3 times the estimate of the B$_{los}$  which then should be considered as an order of magnitude estimate only. For two positions the sign of the line-of-sight component of the vector B is negative.

\subsection{Other objects}

Because of high noise, the observed Stokes \textit{V} signal (see Fig. \ref{fieldRWLMI} and \ref{fieldRYDRA}) allows us to only derive an upper limit of the magnetic field $B_{los}$ along the line-of-sight for the two other AGB stars. We find 14.2 mG and 3.8 mG for RY Dra and RW LMi, respectively (see Table \ref{result}).

For the PPN AFGL618 the Stokes \textit{V} signal is detected above the instrumental contribution and the noise (see Fig. \ref{magnetic field AFGL618} and Table \ref{result}). The magnetic field B$_{los}$ is then estimated to be 6.0 mG for this object. But again, considering the error found, we can only say that we have obtained an order of magnitude estimate of the magnetic field strength B$_{los}$. Concerning the young  PN NGC7027 (see Fig. \ref{fieldNGC7027}), an upper limit of  B$_{los}$ is estimated to 8 mG.


\begin{table}
\centering
\caption{For each object in our sample we give the estimated magnetic field strength on the line-of-sight B$_{los}$, its uncertainty, the strength of the magnetic field  extrapolated (following a $1/r$ law) at one stellar radius (see Sect. 6) B$_{r_{*}}$, and the instrumental contribution parameters $C_1$ and $C_2$.}
\begin{tabular}{lccccc}
\hline
 Object  & $B_{los}$  &  $\delta B_{los}$ & $B_{r_{*}}$ & $\mid C_1 \mid$ & $C_2$ \\ 
 & [mG] & [mG] & [G]  & $\times 10^{-3}$ & [Hz]  \\
\hline
RW LMi &  $ \leq 3.8$ &   &$ \leq 4.4$  & $1.05$ & 9200 \\
RY Dra & $ \leq 14.2 $ &  & $\leq 4.8$ & $1.15$ & 17500\\
AFGL618 &  6.0 & 6.0 & 67.5  & $1.05$ & 2750 \\
NGC7027 & $\leq 8.0$ &  & $\leq 3.1 \times 10^{5}$  & $1.1$ & 1360  \\
\hline
\end{tabular}
\label{result}
\end{table}


\section{Discussion}
\label{sec:discussion}

\subsection{Distribution of the magnetic field in IRC+10216}

Considering the symmetry of the CN ring in Fig. \ref{mapIRC} \citep[see][]{lucas1995}, and assuming that the magnetic field should be stronger where the CN material is denser, we should expect similar values of the magnetic field for different pairs of positions: $(-10'', +20'')/(+15'', -15'')$, $(+20'', +16'')/(-18'', -10'')$, and $(+18'', -04'')/(-18'', +10'')$. However, this is not the case. We observe a stronger magnetic field in the northern part of the ring where CN seems to be less dense.  Furthermore, we could also expect a stronger magnetic field for positions $(+18'', -04'')$ and $(-18'', +10'')$ where CN is observed to be more intense, but this is not the case either, even though position $(-18'', +10'')$ overlaps region $(-10'', +20'')$ where B is strong. The non-detection of the Zeeman effect for position $(+0'',+0'')$ is consistent with the CN hole and with oppositely oriented magnetic field vectors in front and near-side (Fig. \ref{mapIRC}).

Several explanations could explain the non-detections in our observations: 1) CN is less abundant at some positions; 2) the magnetic field vectors cancel out when averaged within the beam; 3) the magnetic field distribution is not homogeneous. 

First of all, the total integrated CN intensities for each position show that the CN distribution has slightly changed since the observations of \citet{lucas1995}: the western part of the CN ring is now the weakest one while the emission coming from the northern part is now stronger. Nevertheless, there is no obvious correlation between the field strength B$_{los}$ and CN emission. 

From this Zeeman effect study we have also inferred the sign, i.e. direction, of the line-of-sight component of the magnetic field vector which is negative for positions $(+15'', -15'')$ and $(+20'', +16'')$ and positive for all other positions. As a consequence no obvious direction pattern is observed. We underline that a magnetic field perpendicular to the observed CN ring would cause the same sign everywhere while a toroidal field within the mapped CN ring, with the torus slightly inclined, would produce a characteristic B$_{los}$ distribution: zero at the center and at the minor axis positions while it would be maximum at the major axis positions with sign reversed from one side to the other. 

\citet{menshi2001} modeled the geometrical structure of the envelope within three regions: the inner most dense core with bipolar cavities (outflow) over 60 AU (0$\arcsec$5), a less-dense envelope where molecules are observed, and the outer extended envelope (6$\times 10^5$ AU $\sim 1^\circ$3). The opening angle of the cavities is 36$^{\circ}$ and the viewing angle between the equatorial plane and the line of sight is 40$^{\circ}$. Therefore, we can extrapolate that the less dense northern and southern parts of the CN ring in the \citet{lucas1995} map correspond to the continuity of the cavities. As a consequence, since 1995, the CN ring might have been modified by a change in the cavities or the viewing angle has changed (unlikely). Moreover, the measured strength of the magnetic field B$_{los}$ could depend on the viewing angle of the outflow cavities assuming that magnetic field vectors follow the outflow cavities.
 
\subsection{Comparison with other observations and implications for the magnetic field mechanism}

We tried to verify that our magnetic field estimates are consistent with previous studies to date exclusively dedicated  to O-rich stars. 
We now intend to link these results to other detections in the O-rich stellar environments. Assuming that the magnetic field process does not depend on the chemical type of the star, and knowing that the CN layer for C-rich objects is roughly at the same distance as OH layer for O-rich objects, we can compare the values of B for these two layers. These results (see Tables \ref{result_IRC} and \ref{result}) are compatible with an estimate of the B$_{los}$ field from OH masers observed for instance by \citet{rudnitski2010} or \citet{gonidakis2014}. 


   \begin{figure}
   \includegraphics[scale=0.45, angle=270]{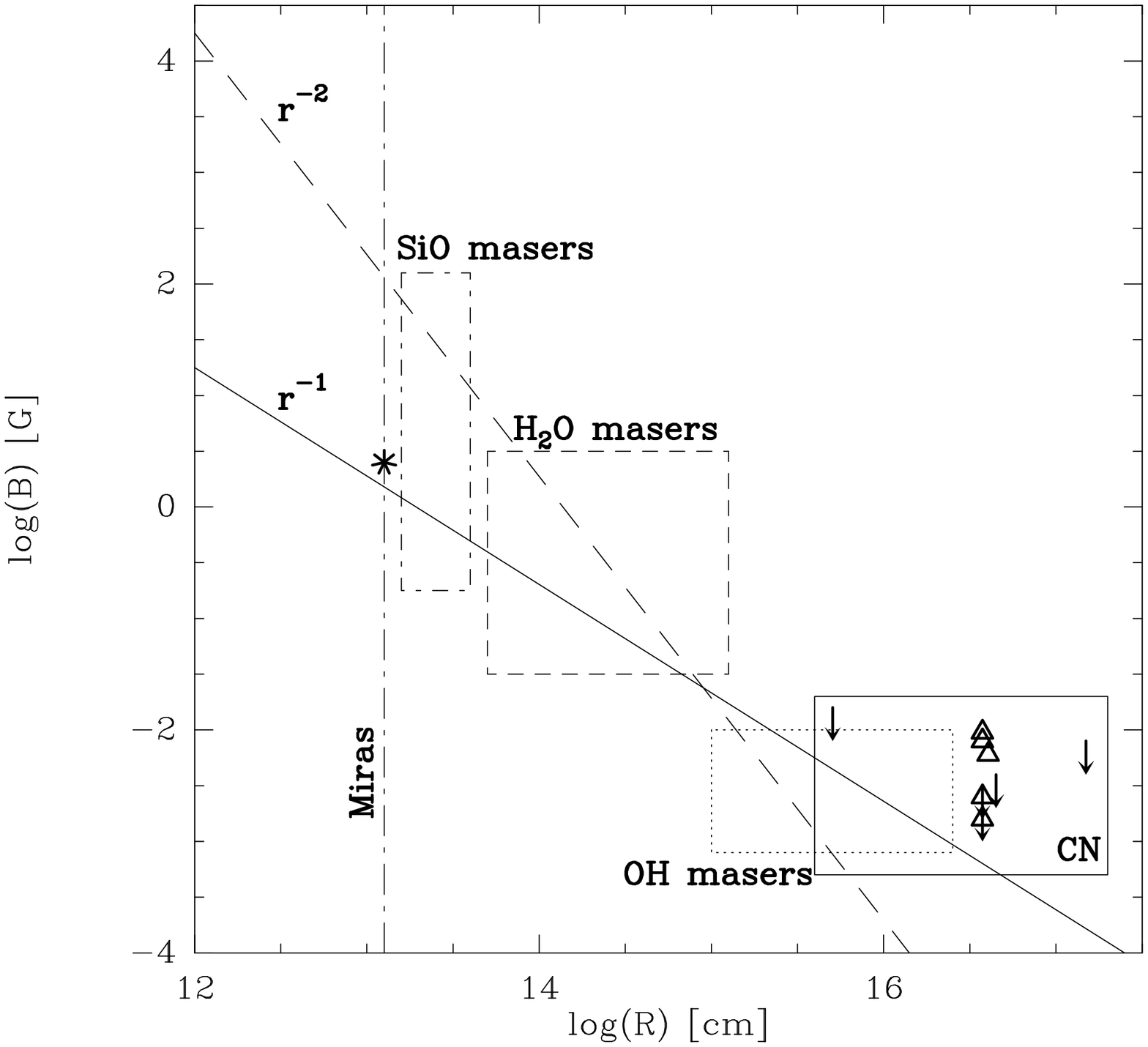}
   \caption{Magnetic field strength as a function of the distance \citep[][]{vlemmings2012}. The different boxes show the range of observed magnetic field strengths derived from observations of SiO masers (\citet{kemball2009}; \citet{herpin2006}), H$_{2}$O masers (\citet{vlemmings2002} and \citet{vlemmings2005}), OH masers (e.g \citet{rudnitski2010}) and CN (this work, triangles for Zeeman detection and arrows for the upper values). The dashed and solid lines indicate an $r^{-2 }$ solar-type and $r^{-1}$ toroidal magnetic field configuration. The vertical dashed line indicates the stellar surface for the Miras, and the star corresponds to the magnetic field measured at the surface of  $\chi$ Cyg by \citet{lebre2014} for this object. }
              \label{graphvlemmings}%
    \end{figure}

Earlier field measurements have shown that the magnetic field strength decreases across the envelope either in $1/r$ or in $1/r^2$  \citep{vlemmings2012}. This is shown in Fig. \ref{graphvlemmings}. Meanwhile, the first estimate of the magnetic field strength (2-3 Gauss) at the surface of a Mira star \citep[$\chi$ Cyg, R$_{\star}=$2 AU,][]{lacour2009} has been obtained by \citet{lebre2014} on \citet[$\chi$ Cyg, R$_{\star}=$2 AU,][]{lacour2009}. Plotted in Fig. \ref{graphvlemmings} it strongly suggests an $1/r$ variation. In order to verify this behavior, we place our B$_{los}$ estimates for all stars in our sample (AGB, PPN and PN) in Fig.\ref{graphvlemmings} which represents the measured magnetic field strength B$_{los}$ versus the radial distance from the center of the star. Our results for C-rich evolved objects confirm that a $1/r$ variation is the most reliable scenario (and definitely exclude a variation in $1/r^3$). All measurements together hence tend to favor a $1/r$ variation of a toroidal magnetic field \citep[as proposed by][]{pascoli1997}.

Adopting this law, we have thus extrapolated the B field to a distance of one stellar radius (B$_{r_{*}}$, see Tables \ref{result_IRC} and \ref{result}). For the AGB stars RY Dra and RW LMi, the upper value for the magnetic field strength is of a few Gauss, while for IRC+10216 the estimated surface magnetic field can be as high as 7.2 G with an average value of 3.8 G. These values are in agreement with the estimate of \citet{lebre2014} for $\chi$ Cyg. For the PPN AFGL618, we derive a surface field of tens of Gauss. In this object, \citet{sabin2014a} have observed a well aligned and organized field along the polar direction (but no toroidal equatorial field) parallel to the major axis of the outflow. Our upper value to B$_{los}$ in the envelope of the PN NGC7027 is compatible with \citet{sabin2007} (a few mG at 3000-4000 AU) or \citet{gomez2009} for the young PN K 3-35 (0.9 mG, based on OH masers). Actually, the estimate by \citet{sabin2007} at 3000-4000 AU from polarimetric SCUBA observations implies that the magnetic field in the CN layer (10000 AU) should be weaker, thus explaining why it is not detected in our study. Moreover, beyond 5000 AU, according to theses authors, there is no sign of an organized field, and, as a consequence, the line-of-sight B$_{los}$ that we measure here might be zero. Nevertheless, the clear correlation observed by these authors between the field orientation and the nebular structure at 3000-4000 AU underlines the importance of the magnetic field in this object. The magnetic field in AFGL618 and NGC7027, with a strength of a few mG, is dominant over the thermal pressure and drives a magnetic outflow launching mechanism.

For the PN the derived upper value of B$_{r_*}$ seems to be far too large compared to previous studies. Actually several studies \citep[e.g.][]{steffen2014} indicate an absence of a strong (kGauss) magnetic field at the PN stellar surface (of the central object), thus in contradiction with our own estimate assuming $1/r$ law. While the stellar radius for AFGL618 \citep[see][]{sanchez2002} is not well constrained, thus the B$_{r_*}$ estimate at the stellar surface remains uncertain, \citet{latter2000} have estimated the radius of the central object of NGC7027 to $3.5\times10^{-4}$ AU. We can then conclude that the magnetic field in the proto-PN and PN does not follow an $1/r$ law or that estimate of the stellar radius for these objects is wrong.

Interpreting IRC+10216 is of course difficult because, as the source is resolved, we have different B estimates. While the non-detection at the central position could be explained by a lower CN abundance, hence a too weak signal, we should expect the same B$_{los}$ field in the two other observed NW and SE positions. On the contrary, only the NW position exhibits a clear Zeeman detection, hence a detectable B$_{los}$ field. This tends to show that the B$_{los}$ field is not homogeneously strong or aligned in the envelope. This result agrees with the study of the magnetic field using molecules CO, SIS, and CS by \citet[][]{girart2012} which suggests that the magnetic field morphology is possibly complex (the positions they studied are within the CN ring). A more detailed interferometric map of the magnetic field is mandatory to conclude, specially as a new spiral structure was recently discovered \citep[][]{cernicharo2015}.

\subsection{Impact on the stellar evolution}
   
The previous section has shown that a $r^{-1}$ decline of the magnetic field across the envelope of O- and C-rich evolved objects is the most likely scenario (except for the PN object in our sample). As a consequence, the magnetic field appears to be toroidal in these evolved objects, and its strength  is expected to be of a few Gauss at the stellar surface. This is in agreement with, for instance, the torus models of \citet{garcia2000}. Nevertheless, our estimates of the magnetic field at the stellar surface, combined with the measurement of \citet{lebre2014} towards the S-type Mira star $\chi$ Cyg, are lower (except for AFGL618) than the prediction from \citet{pascoli2010} of a 10-100 G surface field for an AGB star decreasing as $1/r$. Compared to the field strength required for a toroidal field to launch an outflow via a field pressure gradient \citep[40 G at R$_\star$][]{garcia2005}, the magnetic field, as estimated here, is again too weak at the stellar surface. On the opposite, in the hybrid MHD dust-driven wind model for Mira of \citet{thirumalai2012} the role of a surface field of $\sim$ 4 Gauss is dynamically important in the star's mass loss process. Moreover, \citet{vlemmings2011} has shown, based on the measured magnetic field found in the literature in SiO and H$_2$O, that the magnetic field dominates at and close to the photosphere. This is not the case in the OH$/$CN region, or at least the field energy is comparable to the kinetic energy.

 \section{Conclusion}

Using the polarimeter Xpol with the IRAM 30m, we have made a study of the magnetic field in a small but representative sample of C-rich evolved stars (three AGB stars including the prototypical AGB  IRC+10216, one PPN, and one PN). Thanks to the Zeeman effect in the CN 1-0 transition, we have been able to determine the magnetic field strength using Crutcher's hyperfine lines fitting method. The CN ring observed towards IRC+10216 is well resolved by the 30-m beam and we thus have been able to trace the magnetic field strength in the circumstellar envelope.

This work is the first estimate \citep[apart from a preliminary announcement made by][]{herpin2009} of the magnetic field strength in the circumstellar envelope of C-rich objects. The hyperfine transitions of the CN N=1-0 transition were used to probe the field in these stars and, for the first time, to map the field distribution in the peculiar object IRC +10216.   
   
For AGB stars, we estimate the magnetic field in the CSE to be between 1.5 and 9.5 mG. Previous studies for O-rich evolved stars have show that the magnetic field decreases across the CSE either in 1/r or in 1/r$^2$ (where r is the distance to the star's center), with a preference for 1/r. Considering the magnetic field strength derived in the envelope of the objects studied here and the inferred value $B_{r_{*}}$ at the stellar surface (between 1.1 and 9.5 G), we conclude that the B field varies in 1/r, as expected for a toroidal magnetic field. We stress that such a magnetic field is too weak to launch an outflow via a field pressure gradient. However, a surface magnetic field of few gauss may play an important role in the star's mass loss process. 
  Moreover, our map of IRC+10216 shows that the magnetic field is not homogeneously strong or aligned in the envelope and that the CN morphology might have changed between 1995 and now.
  
   For the central stars of the proto-PN AFGL618 and PN NGC7027, we found B$_{los}$ = 6.0 mG and B$_{los} \leq$ 8.0 mG, respectively, corresponding to an improbably high surface magnetic field of 67 G for AGL618 and an upper limit of 3.4$\times$10$^5$ G for NGC7027 if B varies in $1/r$. For proto-PN and PN, we conclude that the magnetic field might not follow the 1/r law, i.e. something in the stellar evolution between AGB and post-AGB may have changed the field topology. More dedicated polarimetric observations in this class of objects are necessary. 
  
  We have carefully estimated the instrumental contamination in our study. Moreover, considering that we only measure the magnetic field along the line-of-sight, we stress that a no-detection does not necessarily imply that there is no magnetic field. Spectropolarimetric mapping using interferometers like ALMA and NOEMA are required to minimize this problem and to better understand the role of the magnetic field in the evolution of evolved stars.

\begin{acknowledgements}
We would like to thank the anonymous referee for his very useful and constructive comments. We also thank C. Kramer (IRAM-Granada) for his help in the understanding of the leakage issue with XPol.
\end{acknowledgements}
  %
\bibliographystyle{aa}
\bibliography{biblio}

\begin{thebibliography}{77}
\expandafter\ifx\csname natexlab\endcsname\relax\def\natexlab#1{#1}\fi

\bibitem[{{Akashi} {et~al.}(2015){Akashi}, {Sabach}, {Yogev}, \&
  {Soker}}]{akashi2015}
{Akashi}, M., {Sabach}, E., {Yogev}, O., \& {Soker}, N. 2015, \mnras, 453, 2115

\bibitem[{{Alcolea} {et~al.}(2007){Alcolea}, {Neri}, \&
  {Bujarrabal}}]{alcolea2007}
{Alcolea}, J., {Neri}, R., \& {Bujarrabal}, V. 2007, \aap, 468, L41

\bibitem[{{Assaf} {et~al.}(2013){Assaf}, {Diamond}, {Richards}, \&
  {Gray}}]{assaf2013}
{Assaf}, K.~A., {Diamond}, P.~J., {Richards}, A. M.~S., \& {Gray}, M.~D. 2013,
  Monthly Notices of the Royal Astronomical Society, 431, 1077

\bibitem[{{Auri{\`e}re} {et~al.}(2015){Auri{\`e}re}, {Konstantinova-Antova},
  {Charbonnel}, {Wade}, {Tsvetkova}, {Petit}, {Dintrans}, {Drake}, {Decressin},
  {Lagarde}, {Donati}, {Roudier}, {Ligni{\`e}res}, {Schr{\"o}der},
  {Landstreet}, {L{\`e}bre}, {Weiss}, \& {Zahn}}]{auriere2015}
{Auri{\`e}re}, M., {Konstantinova-Antova}, R., {Charbonnel}, C., {et~al.} 2015,
  \aap, 574, A90

\bibitem[{{Bachiller} {et~al.}(1997{\natexlab{a}}){Bachiller}, {Forveille},
  {Huggins}, \& {Cox}}]{bachiller1997b}
{Bachiller}, R., {Forveille}, T., {Huggins}, P.~J., \& {Cox}, P.
  1997{\natexlab{a}}, \aap, 324, 1123

\bibitem[{{Bachiller} {et~al.}(1997{\natexlab{b}}){Bachiller}, {Fuente},
  {Bujarrabal}, {Colomer}, {Loup}, {Omont}, \& {de Jong}}]{bachiller1997}
{Bachiller}, R., {Fuente}, A., {Bujarrabal}, V., {et~al.} 1997{\natexlab{b}},
  \aap, 319, 235

\bibitem[{{Balick} \& {Franck}(2002)}]{balick2002}
{Balick}, B. \& {Franck}, A. 2002, \araa, 40, 439

\bibitem[{{Balick} {et~al.}(2013){Balick}, {Huarte-Espinosa}, {Frank}, {Gomez},
  {Alcolea}, {Corradi}, \& {Vinkovi{\'c}}}]{balick2013}
{Balick}, B., {Huarte-Espinosa}, M., {Frank}, A., {et~al.} 2013, \apj, 772, 20

\bibitem[{{Blackman}(2009)}]{blackman2009}
{Blackman}, E.~G. 2009, in IAU Symposium, Vol. 259, IAU Symposium, ed. K.~G.
  {Strassmeier}, A.~G. {Kosovichev}, \& J.~E. {Beckman}, 35--46

\bibitem[{{Boffin} {et~al.}(2012){Boffin}, {Miszalski}, {Rauch}, {Jones},
  {Corradi}, {Napiwotzki}, {Day-Jones}, \& {K{\"o}ppen}}]{boffin2012}
{Boffin}, H.~M.~J., {Miszalski}, B., {Rauch}, T., {et~al.} 2012, Science, 338,
  773

\bibitem[{{Bujarrabal} \& {Alcolea}(2013)}]{bujarrabal2013}
{Bujarrabal}, V. \& {Alcolea}, J. 2013, \aap, 552, A116

\bibitem[{{Cernicharo} {et~al.}(2015){Cernicharo}, {Marcelino}, {Ag{\'u}ndez},
  \& {Gu{\'e}lin}}]{cernicharo2015}
{Cernicharo}, J., {Marcelino}, N., {Ag{\'u}ndez}, M., \& {Gu{\'e}lin}, M. 2015,
  \aap, 575, A91

\bibitem[{{Chau} {et~al.}(2012){Chau}, {Zhang}, {Nakashima}, {Deguchi}, \&
  {Kwok}}]{chau2012}
{Chau}, W., {Zhang}, Y., {Nakashima}, J.-i., {Deguchi}, S., \& {Kwok}, S. 2012,
  \apj, 760, 66

\bibitem[{{Crutcher} {et~al.}(1996){Crutcher}, {Troland}, {Lazareff}, \&
  {Kaz\`{e}s}}]{crutcher1996}
{Crutcher}, R.~M., {Troland}, T.~H., {Lazareff}, B., \& {Kaz\`{e}s}, I. 1996,
  \apj, 456, 217

\bibitem[{{De Beck} {et~al.}(2010){De Beck}, {Decin}, {de Koter}, {Justtanont},
  {Verhoelst}, {Kemper}, \& {Menten}}]{debeck2010}
{De Beck}, E., {Decin}, L., {de Koter}, A., {et~al.} 2010, \aap, 523, A18

\bibitem[{{De Beck} {et~al.}(2012){De Beck}, {Lombaert}, {Ag{\'u}ndez},
  {Daniel}, {Decin}, {Cernicharo}, {M{\"u}ller}, {Min}, {Royer},
  {Vandenbussche}, {de Koter}, {Waters}, {Groenewegen}, {Barlow}, {Gu{\'e}lin},
  {Kahane}, {Pearson}, {Encrenaz}, {Szczerba}, \& {Schmidt}}]{debeck2012}
{De Beck}, E., {Lombaert}, R., {Ag{\'u}ndez}, M., {et~al.} 2012, \aap, 539,
  A108

\bibitem[{{Decin} {et~al.}(2015){Decin}, {Richards}, {Neufeld}, {Steffen},
  {Melnick}, \& {Lombaert}}]{decin2015}
{Decin}, L., {Richards}, A.~M.~S., {Neufeld}, D., {et~al.} 2015, \aap, 574, A5

\bibitem[{{Desmurs} {et~al.}(2000){Desmurs}, {Bujarrabal}, {Colomer}, \&
  {Alcolea}}]{desmurs2000}
{Desmurs}, J.~F., {Bujarrabal}, V., {Colomer}, F., \& {Alcolea}, J. 2000, \aap,
  360, 189

\bibitem[{{Fong} {et~al.}(2006){Fong}, {Meixner}, {Sutton}, {Zalucha}, \&
  {Welch}}]{fong2006}
{Fong}, D., {Meixner}, M., {Sutton}, E.~C., {Zalucha}, A., \& {Welch}, W.~J.
  2006, \apj, 652, 1626

\bibitem[{{Garc{\'{\i}}a-Segura} \& {L{\'o}pez}(2000)}]{garcia2000}
{Garc{\'{\i}}a-Segura}, G. \& {L{\'o}pez}, J.~A. 2000, \apj, 544, 336

\bibitem[{{Garc{\'{\i}}a-Segura} {et~al.}(2005){Garc{\'{\i}}a-Segura},
  {L{\'o}pez}, \& {Franco}}]{garcia2005}
{Garc{\'{\i}}a-Segura}, G., {L{\'o}pez}, J.~A., \& {Franco}, J. 2005, \apj,
  618, 919

\bibitem[{{Girart} {et~al.}(2012){Girart}, {Patel}, {Vlemmings}, \&
  {Rao}}]{girart2012}
{Girart}, J., {Patel}, N., {Vlemmings}, W., \& {Rao}, R. 2012, \apjl, 751, L20

\bibitem[{{G\'{o}mez} {et~al.}(2009){G\'{o}mez}, {Tafoya}, {Anglada},
  {Miranda}, {Torelles}, {Patel}, \& {Franco-Hern\'{a}ndez}}]{gomez2009}
{G\'{o}mez}, Y., {Tafoya}, D., {Anglada}, G., {et~al.} 2009, \apj, 695, 930

\bibitem[{{Gonidakis} {et~al.}(2014){Gonidakis}, {Chapman}, {Deacon}, \&
  {Green}}]{gonidakis2014}
{Gonidakis}, I., {Chapman}, J.~M., {Deacon}, R.~M., \& {Green}, A.~J. 2014,
  \mnras, 443, 3819

\bibitem[{{Habing}(1996)}]{habing1996}
{Habing}, H.~J. 1996, \aapr, 7, 97

\bibitem[{{Herpin} {et~al.}(2009){Herpin}, {Baudry}, {Josselin}, {Thum}, \&
  {Wiesemeyer}}]{herpin2009}
{Herpin}, F., {Baudry}, A., {Josselin}, E., {Thum}, C., \& {Wiesemeyer}, W.
  2009, International Astronomical Union, 259, 47

\bibitem[{{Herpin} {et~al.}(2006){Herpin}, {Baudry}, {Thum}, {Morris}, \&
  {Wiesemeyer}}]{herpin2006}
{Herpin}, F., {Baudry}, A., {Thum}, C., {Morris}, D., \& {Wiesemeyer}, H. 2006,
  \aap, 450, 667

\bibitem[{{Herpin} {et~al.}(2002){Herpin}, {Goicoechea}, \&
  {Cernicharo}}]{herpin2002}
{Herpin}, F., {Goicoechea}, J.~R., \& {Cernicharo}, J. 2002, \apj, 577, 961

\bibitem[{{H{\"o}fner} {et~al.}(2016){H{\"o}fner}, {Bladh}, {Aringer}, \&
  {Ahuja}}]{hofner2016}
{H{\"o}fner}, S., {Bladh}, S., {Aringer}, B., \& {Ahuja}, R. 2016, \aap, 594,
  A108

\bibitem[{{Houde}(2014)}]{houde2014}
{Houde}, M. 2014, \apj, 795, 27

\bibitem[{{Huang} {et~al.}(2010){Huang}, {Hasegawa}, {Dinh-V-Trung}, {Kwok},
  {Muller}, {Hirano}, {Lim}, {Muthu Mariappan}, \& {Lyo}}]{huang2010}
{Huang}, Z.-Y., {Hasegawa}, T.~I., {Dinh-V-Trung}, {et~al.} 2010, \apj, 722,
  273

\bibitem[{{Huggins} \& {Glassgold}(1982)}]{huggins1982}
{Huggins}, P.~J. \& {Glassgold}, A.~E. 1982, I\apj, 87, 1828

\bibitem[{{Izumiura} \& {Hashimoto}(1999)}]{izumiura1999}
{Izumiura}, H. \& {Hashimoto}, O. 1999, in IAU Symposium, Vol. 191, Asymptotic
  Giant Branch Stars, ed. T.~{Le Bertre}, A.~{Lebre}, \& C.~{Waelkens}, 401

\bibitem[{{Jordan} {et~al.}(2012){Jordan}, {Bagnulo}, {Werner}, \&
  {O\'Toole}}]{jordan2012}
{Jordan}, S., {Bagnulo}, S., {Werner}, K., \& {O\'Toole}, S.~J. 2012, \aap,
  542, A64

\bibitem[{{Josselin} \& {Bachiller}(2003)}]{josselin2003}
{Josselin}, E. \& {Bachiller}, R. 2003, \apjs, 397, 659

\bibitem[{{Kemball} \& {Diamond}(1997)}]{kemball1997}
{Kemball}, A.~J. \& {Diamond}, P.~J. 1997, \apjl, 481, L111

\bibitem[{{Kemball} {et~al.}(2009){Kemball}, {Diamond}, {Gonidakis}, {Mitra},
  {Yim}, {Pan}, \& {Chiang}}]{kemball2009}
{Kemball}, A.~J., {Diamond}, P.~J., {Gonidakis}, I., {et~al.} 2009, \apj, 698,
  1721

\bibitem[{{Kim} {et~al.}(2013){Kim}, {Hsieh}, {Liu}, \& {Taam}}]{kim2013}
{Kim}, H., {Hsieh}, I.-T., {Liu}, S.-Y., \& {Taam}, R.~E. 2013, \apj, 776, 86

\bibitem[{{Kim} {et~al.}(2015){Kim}, {Lee}, {Mauron}, \& {Chu}}]{kim2015}
{Kim}, H., {Lee}, H.-G., {Mauron}, N., \& {Chu}, Y.-H. 2015, \apjl, 804, L10

\bibitem[{{Konstantinova-Antova} {et~al.}(2014){Konstantinova-Antova},
  {Auri{\`e}re}, {Charbonnel}, {Drake}, {Wade}, {Tsvetkova}, {Petit},
  {Schr{\"o}der}, \& {L{\`e}bre}}]{konstantinova2014}
{Konstantinova-Antova}, R., {Auri{\`e}re}, M., {Charbonnel}, C., {et~al.} 2014,
  in IAU Symposium, Vol. 302, IAU Symposium, ed. P.~{Petit}, M.~{Jardine}, \&
  H.~C. {Spruit}, 373--376

\bibitem[{{Kwok} \& {Bignell}(1984)}]{kwok1984}
{Kwok}, S. \& {Bignell}, R.~C. 1984, \apj, 276, 544

\bibitem[{{Lacour} {et~al.}(2009){Lacour}, {Thi{\'e}baut}, {Perrin}, {Meimon},
  {Haubois}, {Pedretti}, {Ridgway}, {Monnier}, {Berger}, {Schuller},
  {Woodruff}, {Poncelet}, {Le Coroller}, {Millan-Gabet}, {Lacasse}, \&
  {Traub}}]{lacour2009}
{Lacour}, S., {Thi{\'e}baut}, E., {Perrin}, G., {et~al.} 2009, \apj, 707, 632

\bibitem[{{Latter} {et~al.}(2000){Latter}, {Dayal}, {Bieging}, {Meakin},
  {Hora}, {Kelly}, \& {Tielens}}]{latter2000}
{Latter}, W.~B., {Dayal}, A., {Bieging}, J.~H., {et~al.} 2000, \apj, 539, 783

\bibitem[{{Leal-Ferreira} {et~al.}(2013){Leal-Ferreira}, {Vlemmiings},
  {Kemball}, \& {Amiri}}]{leal2013}
{Leal-Ferreira}, M.~L., {Vlemmiings}, W. H.~T., {Kemball}, A., \& {Amiri}, N.
  2013, \aap, 554, A134

\bibitem[{{Leal-Ferreira} {et~al.}(2012){Leal-Ferreira}, {Vlemmings},
  {Diamond}, {Kemball}, {Amiri}, \& {Desmurs}}]{leal2012}
{Leal-Ferreira}, M.~L., {Vlemmings}, W.~H.~T., {Diamond}, P.~J., {et~al.} 2012,
  \aap, 540, A42

\bibitem[{{L\`{e}bre} {et~al.}(2014){L\`{e}bre}, {Fabas}, {Gillet}, {Herpin},
  {Konstantinova-Antova}, \& {Petit}}]{lebre2014}
{L\`{e}bre}, A.and~{Auri\`{e}re}, M., {Fabas}, N., {Gillet}, D., {et~al.} 2014,
  \aap, 561, A85

\bibitem[{{Lee} {et~al.}(2013){Lee}, {Sahai}, {S{\'a}nchez Contreras}, {Huang},
  \& {Hao Tay}}]{lee2013b}
{Lee}, C.-F., {Sahai}, R., {S{\'a}nchez Contreras}, C., {Huang}, P.-S., \& {Hao
  Tay}, J.~J. 2013, \apj, 777, 37

\bibitem[{{Leone} {et~al.}(2014){Leone}, {Corradi}, {Mart{\'{\i}}nez
  Gonz{\'a}lez}, {Asensio Ramos}, \& {Manso Sainz}}]{leone2014}
{Leone}, F., {Corradi}, R.~L.~M., {Mart{\'{\i}}nez Gonz{\'a}lez}, M.~J.,
  {Asensio Ramos}, A., \& {Manso Sainz}, R. 2014, \aap, 563, A43

\bibitem[{{Lindqvist} {et~al.}(2000){Lindqvist}, {Sch{\"o}ier}, {Lucas}, \&
  {Olofsson}}]{lindqvist2000}
{Lindqvist}, M., {Sch{\"o}ier}, F.~L., {Lucas}, R., \& {Olofsson}, H. 2000,
  \aap, 361, 1036

\bibitem[{{Lucas} {et~al.}(1995){Lucas}, {Guelin}, {Kahane}, {Audinos}, \&
  {Cernicharo}}]{lucas1995}
{Lucas}, R., {Guelin}, M., {Kahane}, C., {Audinos}, P., \& {Cernicharo}, J.
  1995, Astrophyscis and Space Science, 224, 293

\bibitem[{{Masson}(1989)}]{masson1989}
{Masson}, C.~R. 1989, \apj, 336, 294

\bibitem[{{Men'shchikov} {et~al.}(2001){Men'shchikov}, {Balega}, {Bl{\"o}cker},
  {Osterbart}, \& {Weigelt}}]{menshi2001}
{Men'shchikov}, A.~B., {Balega}, Y., {Bl{\"o}cker}, T., {Osterbart}, R., \&
  {Weigelt}, G. 2001, \aap, 368, 497

\bibitem[{{Nordhaus} {et~al.}(2007){Nordhaus}, {Blackman}, \&
  {Frank}}]{nordhaus2007}
{Nordhaus}, J., {Blackman}, E.~G., \& {Frank}, A. 2007, \mnras, 376, 599

\bibitem[{{Nyman} {et~al.}(1998){Nyman}, {Hall}, \& {Olofsson}}]{nyman1998}
{Nyman}, L.-A., {Hall}, P.~J., \& {Olofsson}, H. 1998, \aaps, 127, 185

\bibitem[{{Pascoli}(1997)}]{pascoli1997}
{Pascoli}, G. 1997, \apj, 489, 946

\bibitem[{{Pascoli} \& {Lahoche}(2008)}]{pascoli2008}
{Pascoli}, G. \& {Lahoche}, L. 2008, \pasp, 120, 1267

\bibitem[{{Pascoli} \& {Lahoche}(2010)}]{pascoli2010}
{Pascoli}, G. \& {Lahoche}, L. 2010, \pasp, 122, 1334

\bibitem[{{Ragland} {et~al.}(2006){Ragland}, {Traub}, {Berger}, {Danchi},
  {Monnier}, {Willson}, {Carleton}, {Lacasse}, {Millan-Gabet}, {Pedretti},
  {Schloerb}, {Cotton}, {Townes}, {Brewer}, {Haguenauer}, {Kern}, {Labeye},
  {Malbet}, {Malin}, {Pearlman}, {Perraut}, {Souccar}, \&
  {Wallace}}]{ragland2006}
{Ragland}, S., {Traub}, W.~A., {Berger}, J.-P., {et~al.} 2006, \apj, 652, 650

\bibitem[{{Ramstedt} \& {Olofsson}(2014)}]{ramstedt2014}
{Ramstedt}, S. \& {Olofsson}, H. 2014, \aap, 566, A145

\bibitem[{{Rudnitski} {et~al.}(2010){Rudnitski}, {Pashchenko}, \&
  {Colom}}]{rudnitski2010}
{Rudnitski}, G.~M., {Pashchenko}, M.~I., \& {Colom}, P. 2010, Astronomy
  Reports, 54, 400

\bibitem[{{Sabin} {et~al.}(2015){Sabin}, {Wade}, \& {L{\`e}bre}}]{sabin2015}
{Sabin}, L., {Wade}, G.~A., \& {L{\`e}bre}, A. 2015, \mnras, 446, 1988

\bibitem[{{Sabin} {et~al.}(2014){Sabin}, {Zhang}, {Zijlstra}, {Patel},
  {V{\'a}zquez}, {Zauderer}, {Contreras}, \& {Guill{\'e}n}}]{sabin2014a}
{Sabin}, L., {Zhang}, Q., {Zijlstra}, A.~A., {et~al.} 2014, \mnras, 438, 1794

\bibitem[{{Sabin} {et~al.}(2007){Sabin}, {Zijlstra}, \& {Greaves}}]{sabin2007}
{Sabin}, L., {Zijlstra}, A.~A., \& {Greaves}, J.~S. 2007, \mnras, 376, 378

\bibitem[{{S{\'a}nchez Contreras} {et~al.}(2002){S{\'a}nchez Contreras},
  {Sahai}, \& {Gil de Paz}}]{sanchez2002}
{S{\'a}nchez Contreras}, C., {Sahai}, R., \& {Gil de Paz}, A. 2002, \apj, 578,
  269

\bibitem[{{Schmidt} {et~al.}(2002){Schmidt}, {Hines}, \& {Swift}}]{schmidt2002}
{Schmidt}, G.~D., {Hines}, D.~C., \& {Swift}, S. 2002, \apj, 576, 429

\bibitem[{{Skinner} {et~al.}(1998){Skinner}, {Meixner}, \&
  {Bobrowsky}}]{skinner1998}
{Skinner}, C.~J., {Meixner}, M., \& {Bobrowsky}, M. 1998, \mnras, 300, L29

\bibitem[{{Soker}(2000)}]{soker2000}
{Soker}, N. 2000, \apj, 540, 436

\bibitem[{{Steffen} {et~al.}(2014){Steffen}, {Hubrig}, {Todt}, {Sch{\"o}ller},
  {Hamann}, {Sandin}, \& {Sch{\"o}nberner}}]{steffen2014}
{Steffen}, M., {Hubrig}, S., {Todt}, H., {et~al.} 2014, \aap, 570, A88

\bibitem[{{Thirumalai} \& {Heyl}(2012)}]{thirumalai2012}
{Thirumalai}, A. \& {Heyl}, J.~S. 2012, \mnras, 422, 1272

\bibitem[{{Thum} {et~al.}(2008){Thum}, {Wiesemeyer}, {Paubert}, {Navarro}, \&
  {Morris}}]{thum2008}
{Thum}, C., {Wiesemeyer}, H., {Paubert}, G., {Navarro}, S., \& {Morris}, D.
  2008, \pasp, 120, 777

\bibitem[{{Tuthill} {et~al.}(2000){Tuthill}, {Monnier}, {Danchi}, \&
  {Lopez}}]{tuthill2000}
{Tuthill}, P.~G., {Monnier}, J.~D., {Danchi}, W.~C., \& {Lopez}, B. 2000, \apj,
  543, 284

\bibitem[{{Vlemmings}(2011)}]{vlemmings2011}
{Vlemmings}, W.~H.~T. 2011, in Asymmetric Planetary Nebulae 5 Conference, 89

\bibitem[{{Vlemmings}(2012)}]{vlemmings2012}
{Vlemmings}, W.~H.~T. 2012, in IAU Symposium, Vol. 287, Cosmic Masers - from OH
  to H0, ed. R.~S. {Booth}, W.~H.~T. {Vlemmings}, \& E.~M.~L. {Humphreys},
  31--40

\bibitem[{{Vlemmings} {et~al.}(2001){Vlemmings}, {Diamond}, \& {van
  Langevelde}}]{vlemmings2001}
{Vlemmings}, W. H.~T., {Diamond}, P., \& {van Langevelde}, H.~J. 2001, \aap,
  375, 6L1

\bibitem[{{Vlemmings} {et~al.}(2002){Vlemmings}, {Diamond}, \& {van
  Langevelde}}]{vlemmings2002}
{Vlemmings}, W. H.~T., {Diamond}, P., \& {van Langevelde}, H.~J. 2002, \aap,
  394, 589

\bibitem[{{Vlemmings} {et~al.}(2005){Vlemmings}, {van Langevelde}, \&
  {Diamond}}]{vlemmings2005}
{Vlemmings}, W. H.~T., {van Langevelde}, H.~J., \& {Diamond}, P. 2005, \aap,
  434, 1029

\bibitem[{{Western} \& {Watson}(1983)}]{western1983}
{Western}, L.~R. \& {Watson}, W.~D. 1983, \apj, 275, 195

\end{thebibliography}

\newpage

 \begin{appendix}
 \section{Figures}
\begin{figure}
   \centering
   \includegraphics[width=190pt, angle=270]{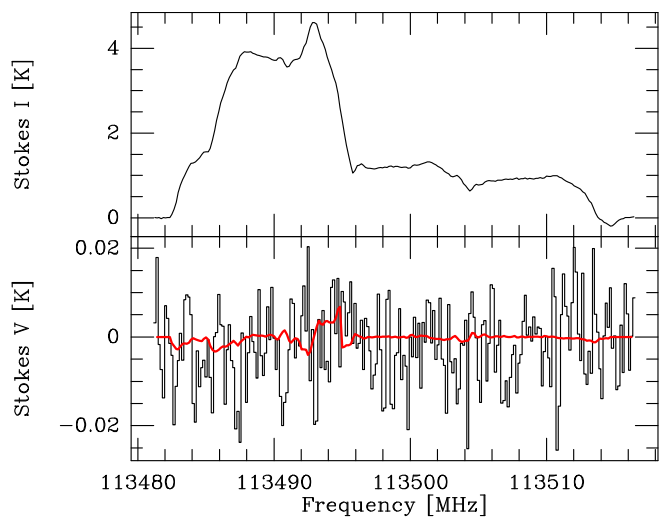}
   \caption{ IRC+10216: observations of November 2006 for the position (+15\arcsec, -15\arcsec) and the CN transition (1, 3/2) $\rightarrow$ (0,1/2). \textit{Top}: Stokes \textit{I} spectrum. \textit{Bottom}: spectra and least-squares fits to \textit{V} are shown in black and red, respectively.}
              \label{magnetic field IRC+15-15}%
    \end{figure}

\begin{figure}
   \includegraphics[width=200pt, angle=270]{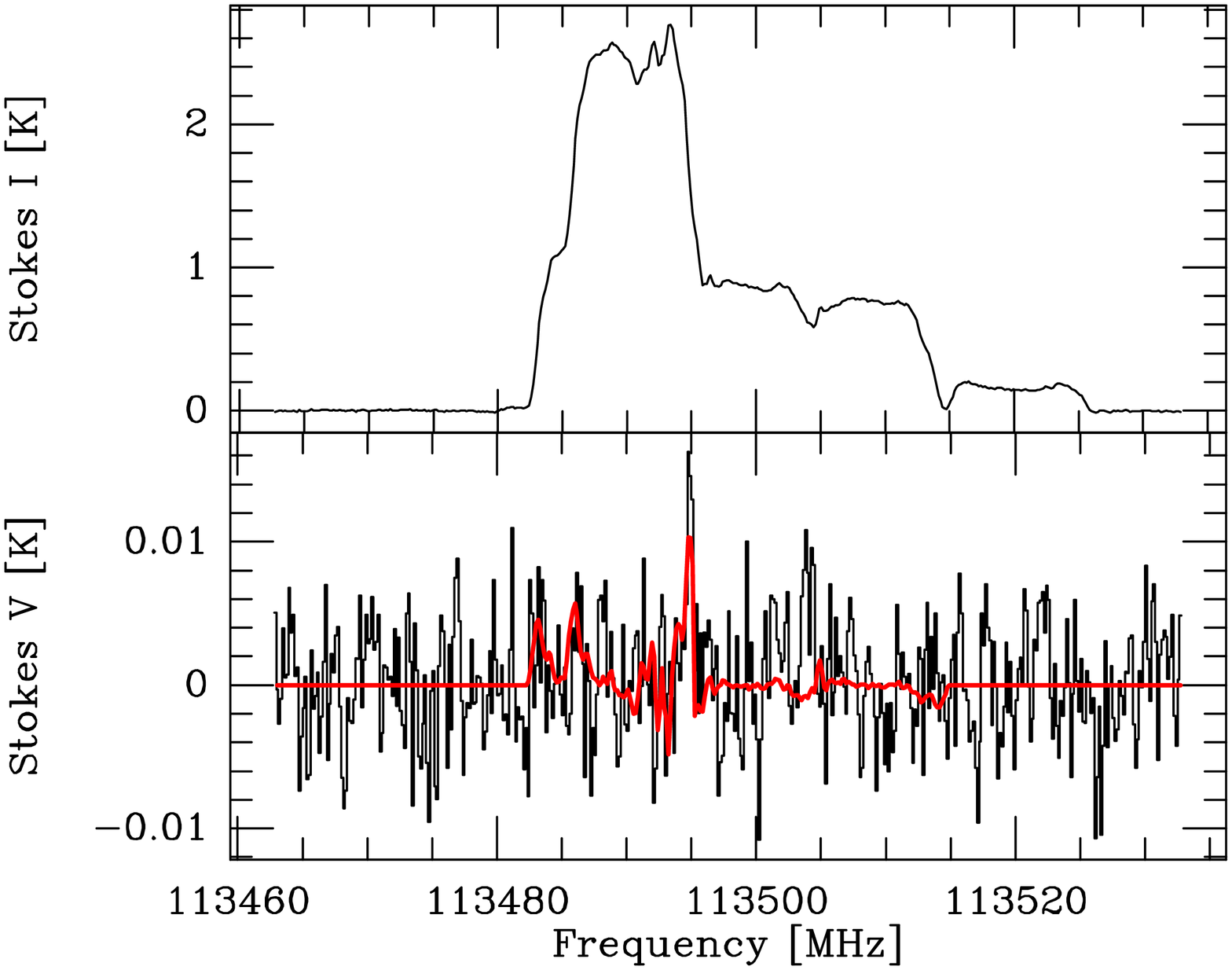}
   \caption{As in Fig. \ref{magnetic field IRC+15-15} but for the observations of March and June 2016 toward position (-18\arcsec, +10\arcsec) of IRC+10216.}
              \label{fieldRC2}%
    \end{figure}

\begin{figure}
   \includegraphics[width=200pt, angle=270]{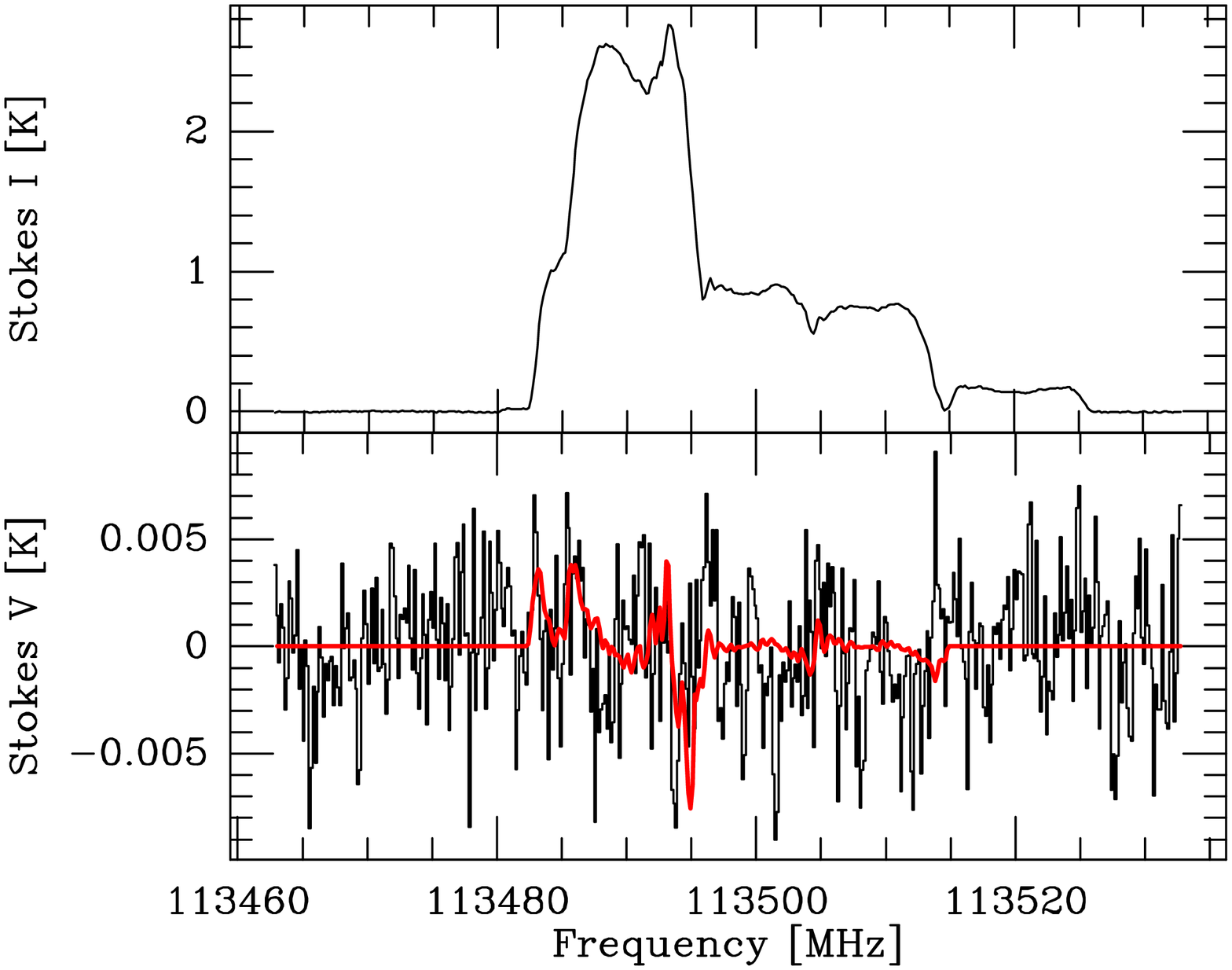}
   \caption{As in Fig. \ref{magnetic field IRC+15-15} but for the observations of March and June 2016 toward position (-18\arcsec, -10\arcsec) of IRC+10216.}
              \label{fieldRC3}%
    \end{figure}

\begin{figure}
   \includegraphics[width=200pt, angle=270]{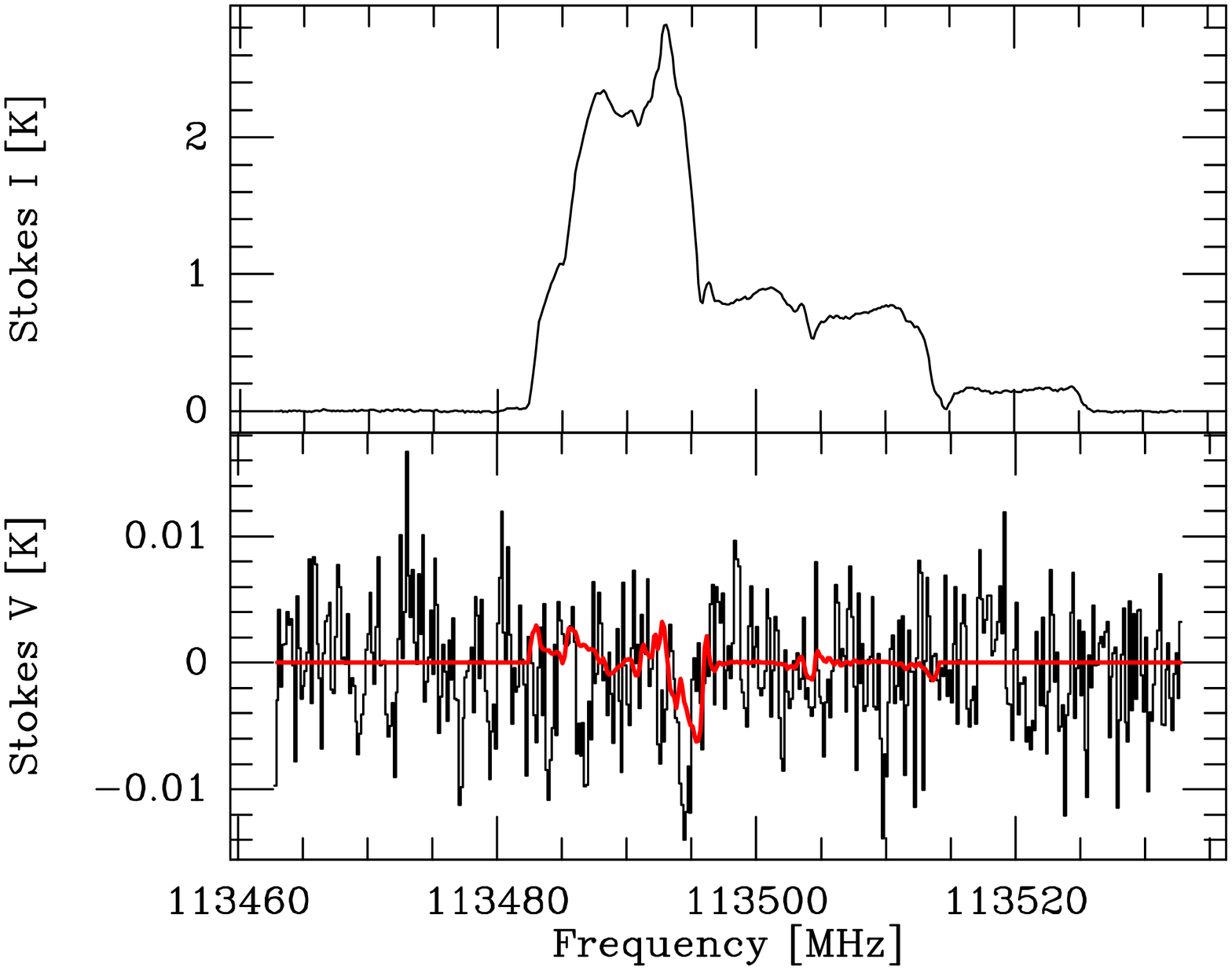}
   \caption{As in Fig. \ref{magnetic field IRC+15-15} but for the observations of March 2016 toward position (+18\arcsec, -04\arcsec)  of IRC+10216.}
              \label{fieldRC5}%
    \end{figure}

\begin{figure}
   \includegraphics[width=200pt, angle=270]{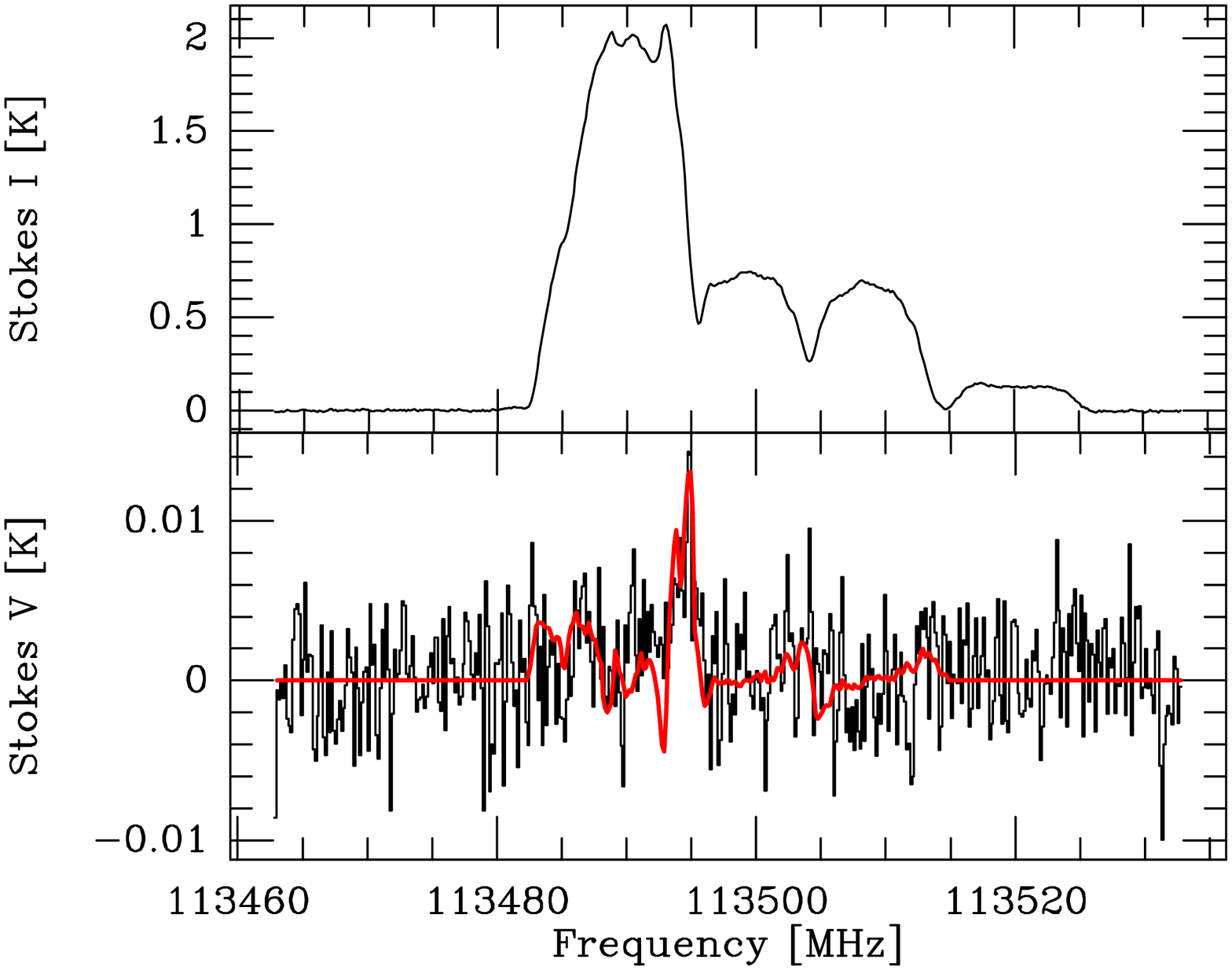}
   \caption{As in Fig. \ref{magnetic field IRC+15-15} but for the observations of March and June 2016 toward position (+20\arcsec, +16\arcsec) of IRC+10216.}
              \label{fieldRC6}%
    \end{figure}

\begin{figure}
   \includegraphics[width=200pt, angle=270]{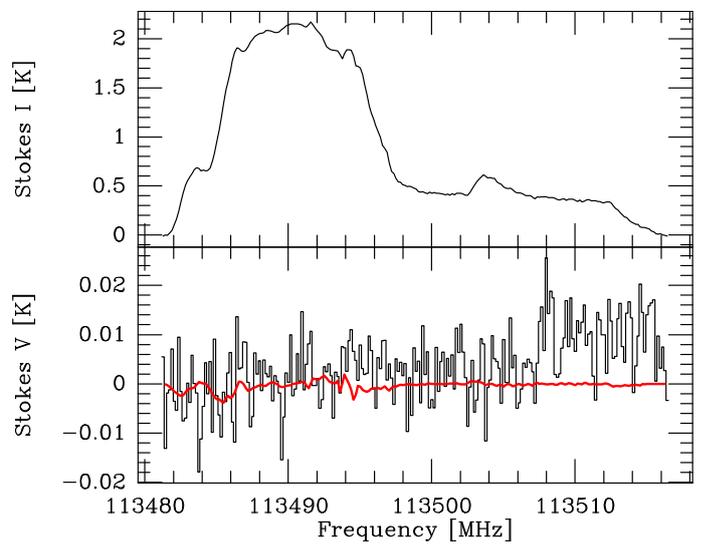}
   \caption{As in Fig. \ref{magnetic field IRC+15-15} but for the observations of November 2006 toward RW LMi.}
              \label{fieldRWLMI}%
    \end{figure}

\begin{figure}
   \includegraphics[width=200pt, angle=270]{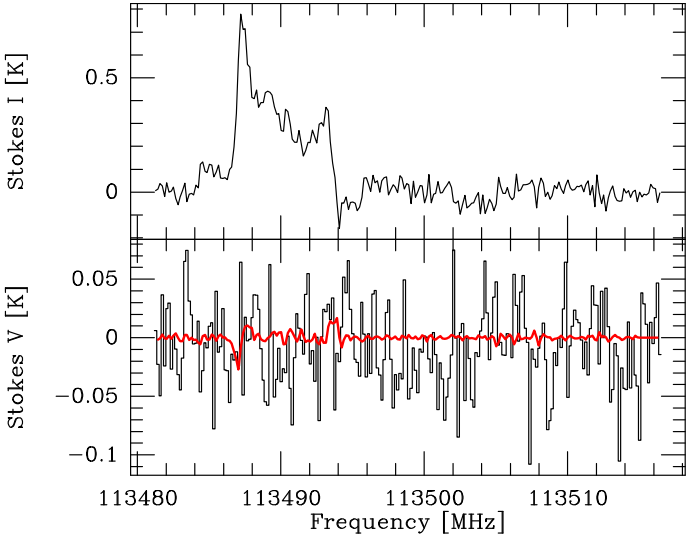}
   \caption{As in Fig. \ref{magnetic field IRC+15-15} but for the observations of November 2006 toward RY Dra.}
              \label{fieldRYDRA}%
    \end{figure}

\begin{figure}
   \includegraphics[width=195pt, angle=270]{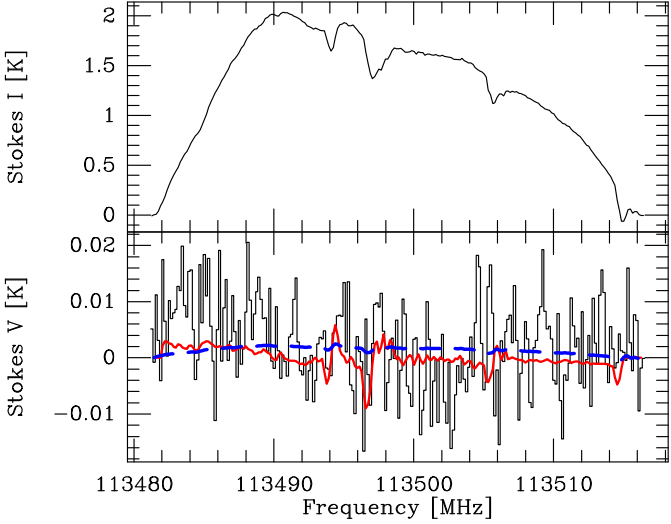}
   \caption{AFGL618: observations of November 2006. CN (1, 3/2) $\rightarrow$ (0, 1/2) Stokes \textit{I} (\textit{Top}) and \textit{V} (\textit{Bottom}) spectra. Spectra and least-squares fits to \textit{V} are shown in black and red, respectively. The instrumental \textit{V} contribution is plotted in blue ($C_3=0$).}
              \label{magnetic field AFGL618}%
    \end{figure}

\begin{figure}
   \includegraphics[width=200pt, angle=270]{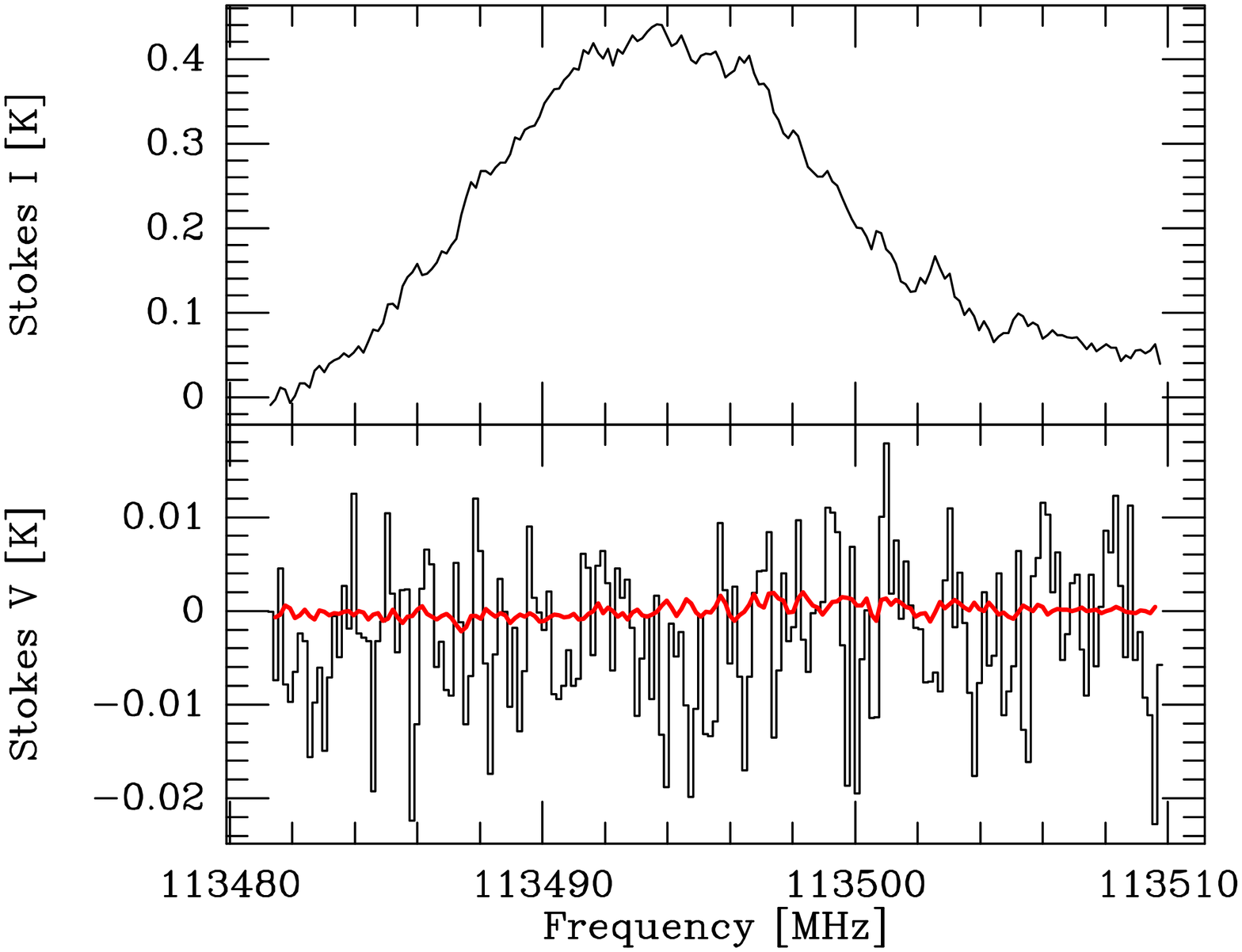}
   \caption{As in Fig. \ref{magnetic field IRC+15-15} but for the observations of November 2006 toward NCC7027.}
              \label{fieldNGC7027}%
    \end{figure}

Figures \ref{magnetic field IRC+15-15} to \ref{fieldNGC7027} show the stokes \textit{I} and \textit{V} spectra for all stars in our sample.


%
%


%
%

\section{Power patterns in Stokes \textit{I} and \textit{V}}
\label{appendixB}
\begin{figure}
   \includegraphics[width=280pt]{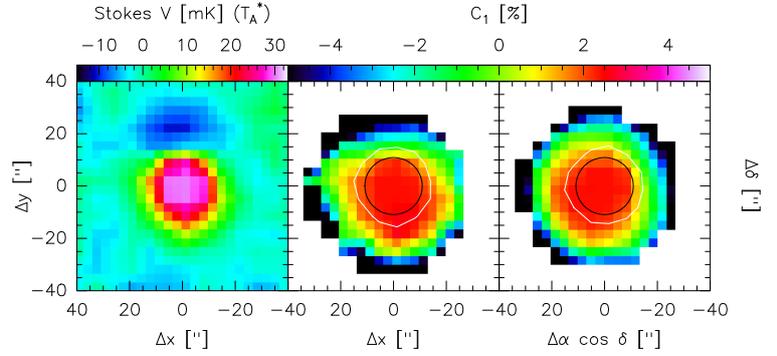}
   \caption{Stokes \textit{V} power patterns, as measured on Uranus.{\bf Left:} Nasmyth reference frame. The color scale
indicates the antenna temperature (in mK). {\bf Center:} Same for $C_1$, with color scale in \%.
The white contour indicates the measured half-maximum contour of the Stokes I beam, at 91.5~GHz. The black contour
is the corresponding contour at the CN$(1-0)$ frequency. The map is limited by a $S/N \sim 5$ cutoff. {\bf Right:} Averaged $C_1$ map in astronomical
coordinates. For details see text.
     }
              \label{Uranus}%
    \end{figure}

As mentioned in section \ref{instr_cont}, the parameter $C_1$ is related to the leakage of Stokes \textit{I} into Stokes \textit{V}, in such a way that the corresponding power pattern can be described by a scaled copy of the Stokes \textit{I} beam. In general, the power patterns are measured by observing a sufficiently strong, unpolarized and unresolved continuum source. 
When EMIR's orthomode transducers were commissioned, a serie of observations of Uranus was made in December 2015 at various elevations. While on the optical axis a value of 2.5~\% was measured for $C_1$, it increases off the optical axis (Fig.~\ref{Uranus}).

In the following, the indexing of the power patterns follows the nomenclature of the M{\"u}ller matrices, e.g.,
IV describes the conversion from Stokes \textit{I} into Stokes \textit{V} in the telescope's Nasmyth cabin, and $B_{\rm IV}$ is
the corresponding power pattern. The observed Stokes \textit{V} then becomes
\begin{eqnarray}
V_{\rm obs} & = & V_{\rm int} \ast B_{\rm VV} + I_{\rm int} \ast B_{\rm IV}\, \nonumber \\
I_{\rm obs} & = & I_{\rm int} \ast B_{\rm II}\,,
\label{eq:ip}
\end{eqnarray}
where $I_{\rm int}$ and $V_{\rm int}$ are the intrinsic brightness distributions in Stokes \textit{I} and \textit{V}, respectively, and $I_{\rm obs}$ and $V_{\rm obs}$ are the observed flux densities after convolution with the respective power patterns of the antenna. 
Since the beams $B_{\rm II}$ and $B_{\rm VV}$ are dominated by the aperture of the telescope, and the illumination of its subreflector by the receiver, we assume them to be equal. As already mentioned, the power distribution leakage of Stokes \textit{I} into Stokes \textit{V}, $B_{\rm IV}$, has been measured (Fig.~\ref{Uranus}). The spatial distribution of the observed fractional circular polarization is then given by
\begin{equation}
\left(\frac{V_{\rm obs}}{I_{\rm obs}}\right) = C_3 \frac{d \ln I}{d\nu} + C_1 = \frac{V_{\rm int} \ast B_{\rm VV} + I_{\rm int} \ast B_{\rm IV}}{I_{\rm int} \ast B_{\rm II}}\,.
\label{eq:c1_c3}
\end{equation}

where, as discussed in section \ref{instr_cont}, we can neglect the term with $C_2$. The only quantity which varies with time is $B_{\rm IV}$. Because it is not entirely axially symmetric, but displays an asymmetric sidelobe which is fixed in the Nasmyth reference frame, the equivalent power pattern $B_{\rm IV}$ is smeared by the parallactic rotation. This is demonstrated in Fig.~\ref{Uranus}.
Eq.~\ref{eq:c1_c3} shows that applying Crutcher's method to time-averaged spectra, with a correspondingly weighted averaged $B_{\rm IV}$ is equivalent to applying the method to data subsets, at various elevations and parallactic angles.
The Stokes \textit{V} spectra of these subsets should then be corrected individually. However, the pattern method will be fraught with uncertainties, because the lower sensitivity in subsets of the spectra may introduce artefacts into the determination of the $C$ coefficients. We therefore prefer to use the method described in Sect. \ref{sec:discussion}. The value of $C_1$ then depends on the emission picked up from e.g. the CN(1-0) emission shell of IRC+10216 which is described by the aforementioned power pattern $B_{\rm IV}$. This explains why the $C_1$ coefficients vary from position to position (cf. Table~\ref{result_IRC}). While the average $C_1$ within a $5\sigma$ cutoff radius in Stokes \textit{I} is $-0.3$~\%, the average between the half-power contour of the Stokes \textit{I} beam and the same cutoff is $-0.5$~\%, varying between $-0.4$~\% and $-1.4$~\% within the four quadrants of this area.
These values are reasonably close, in absolute value {\it and} sign, to those given in Table~\ref{result_IRC}. In practice, the exact values depend on where the dominating pickup of Stokes \textit{I} by the power pattern $B_{\rm IV}$ is located. A more quantitative analysis would only be possible with a high resolution ($\sim 1$\arcsec) map of the CN(1-0) emission in Stokes \textit{I}.
 
%

 
 

 \end{appendix}

\end{document}